\documentclass[aps,prb, twocolumn,superscriptaddress, noeprint]{revtex4-2}
\usepackage{bbold}
\usepackage{comment}
\usepackage[colorlinks, citecolor=red]{hyperref}
\usepackage{hyperref}
\usepackage[utf8]{inputenc}
\usepackage[T1]{fontenc}    %
\usepackage[english]{babel}
\usepackage[pdftex]{graphicx}
\usepackage{amsmath,amssymb,amsthm,bbm, mathtools} %
\usepackage{mathrsfs}
\usepackage{xcolor}
\usepackage[normalem]{ulem}

\DeclareMathSymbol{\shortminus}{\mathbin}{AMSa}{"39}
\renewcommand{\vec}{\boldsymbol}

\begin{document}

\title{High-fidelity two-qubit gates of hybrid superconducting-semiconducting \\ singlet-triplet qubits}

\author{Maria Spethmann}
\email{maria.spethmann@unibas.ch}
\affiliation{Department of Physics, University of Basel, 4056 Basel, Switzerland}
\affiliation{RIKEN, Center for Emergent Matter Science (CEMS), Wako-shi, Saitama 351-0198, Japan}
\author{Stefano Bosco}
\affiliation{Department of Physics, University of Basel, 4056 Basel, Switzerland}
\author{Andrea Hofmann}
\affiliation{Department of Physics, University of Basel, 4056 Basel, Switzerland}
\author{Jelena Klinovaja}
\affiliation{Department of Physics, University of Basel, 4056 Basel, Switzerland}
\author{Daniel Loss}
\affiliation{Department of Physics, University of Basel, 4056 Basel, Switzerland}
\affiliation{RIKEN, Center for Emergent Matter Science (CEMS), Wako-shi, Saitama 351-0198, Japan}

\begin{abstract}
Hybrid systems comprising superconducting and semiconducting materials are promising architectures for quantum computing. 
Superconductors induce long-range interactions between the spin degrees of freedom of semiconducting quantum dots. These interactions are widely anisotropic when the semiconductor material has strong spin-orbit interactions. 
We show that this anisotropy is tunable and enables fast and high-fidelity two-qubit gates between singlet-triplet (ST) spin qubits. Our design is immune to leakage of the quantum information into noncomputational states and removes always-on interactions between the qubits, thus resolving key open challenges for these architectures. 
Our ST qubits do not require additional technologically demanding components nor fine-tuning of parameters. They operate at low magnetic fields of a few millitesla and are fully compatible with superconductors. 
By suppressing systematic errors in realistic devices, we estimate infidelities below $10^{-3}$, which could pave the way toward large-scale hybrid superconducting-semiconducting quantum processors.

\end{abstract}

\maketitle

\begin{figure}[t]
	\includegraphics[width=0.49\textwidth]{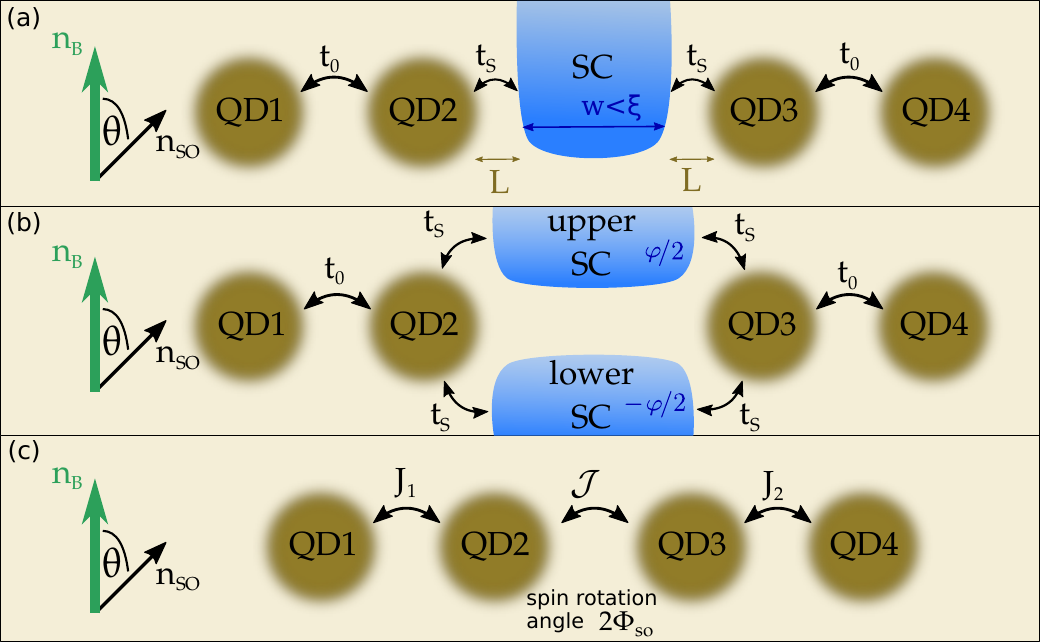}
\caption{
Schematics of coupled hybrid ST qubits. Two ST qubits, each comprising a double quantum dot, interact (a) via one superconductor  or (b) via a Josephson junction. (c) These setups are effectively equivalent to two exchange-coupled double quantum dots with fully tunable interactions.
Strong SOI induces spin rotations of an angle $\Phi_\text{so}$ around an axis $\vec{n}_\text{so}$ and yield a large asymmetry of exchange coupling depending on the angle $ \theta$ between $\vec{n}_\text{so}$ and the Zeeman vector $\vec{n}_B$, which is determined by the direction of the applied magnetic field $\vec{B}$. %
\label{pic:13209801293}}

\end{figure}

\section{Introduction }

Hybrid systems comprising superconductors and semiconductors are the workhorse of modern quantum technology, with applications in low-power electronics~\cite{McCaughan2019} and in neuromorphic~\cite{PhysRevApplied.7.034013} and quantum computing~\cite{Burkard2020,PhysRevLett.90.087003}.
Front-runner quantum bits (qubits) are encoded in the spin of particles confined in semiconducting quantum dots \cite{Loss1998,Gilbert2023, Burkard2021,Stano2021,Philips2022} or in collective modes of superconducting devices~\cite{RevModPhys.93.025005,Arute2019}. Spin qubits are compact but challenging to address, while superconducting qubits are bulky but easy to couple.
By combining the best properties of each architecture, hybrid qubits \cite{Hays2021,Chtchelkatchev2003,Pita-Vidal2022, bargerbos2022spectroscopy,Leijnse2013}
could outperform the state of the art and pave the way toward large-scale quantum processors.

A key ingredient for effective hybrid systems is spin-orbit interaction (SOI). In Josephson junctions, SOI induces %
a spin-dependent supercurrent that is critical to manipulate and read out Andreev spin qubits \cite{Chtchelkatchev2003, Hays2021, Pita-Vidal2022, bargerbos2022spectroscopy}. Topological encoding of quantum information in Majorana bound states~\cite{RevModPhys.83.1057,Leijnse_2012,doi:10.1146/annurev-conmatphys-030212-184337,doi:10.7566/JPSJ.85.072001,doi:10.1063/5.0055997,Hoffman2016}
and long-range entanglement of distant spins~\cite{yu2022strong,PhysRevLett.129.066801,PhysRevB.107.L041303,PhysRevResearch.3.013194,PhysRevX.12.021026} also crucially require strong effective SOIs. 
SOIs are exceptionally large in narrow-gap semiconducting nanowires~\cite{Nadj-Perge2010,PhysRevLett.110.066806,Liang2012,PhysRevLett.122.187702,PhysRevB.91.201413} and in nanostructures where the charge carriers are holes rather than electrons~\cite{Froning2021b,Wang2022b,Watzinger2018,Maurand2016,PhysRevResearch.3.013081,PhysRevB.84.195314,PhysRevB.97.235422,PRXQuantum.2.010348,Wang2021,Adelsberger2022enhanced,Adelsberger2022hole}. Hole gases in planar germanium (Ge) heterostructures \cite{Scappucci2021,Hendrickx2021,Hendrickx2020,Hendrickx20201,PhysRevB.104.115425,PhysRevB.103.125201} are particularly appealing because of their compatibility with superconducting materials, the possibility of engineered proximitized superconductivity~\cite{PhysRevResearch.3.L022005,tosato2022hard,Vigneau2019,Hendrickx2018,PhysRevB.107.035435,PhysRevB.84.104526}, and sweet spots to reduce noise \cite{PhysRevB.104.115425, Bosco2021fully, Bosco2022hole}. Recent experiments with Ge \cite{Jirovec2021,Jirovec2022} demonstrated operations of singlet-triplet (ST) spin qubits, encoding quantum information in the zero-spin subspace of two coupled quantum dots \cite{Levy2002}, at millitesla magnetic fields~\cite{Jirovec2021}. These fields are compatible with current superconducting devices, opening various opportunities for hybrid systems in a potentially nuclear spin free material such as Ge.  

In this work, we discuss a robust implementation of high-fidelity two-qubit gates between distant ST qubits \cite{Levy2002, Petta2005, Jirovec2021,Jirovec2022, liles2023, Fedele2021,Cerfontaine2020} in hybrid systems \cite{Leijnse2013}; see Fig.~\ref{pic:13209801293}. By taking full advantage of both the large SOI in the material and the long-range spin-spin correlations induced by the superconductor~\cite{Choi2000,Wang2022,Bordoloi2022}, our design overcomes the fundamental limitations of current two-qubit gates between ST qubits, namely leakage to noncomputational states~\cite{Levy2002, Klinovaja2012, Li2012, PhysRevB.91.085419} and crosstalk~\cite{Cerfontaine20201,PhysRevB.97.045431} caused by always-on residual interactions between dots.

We find that leakage is naturally suppressed by the large SOI in hole systems, yielding a tunable anisotropy in the exchange interactions of quantum dots~\cite{geyer2022two}, and crosstalk  vanishes by utilizing the phase response of the supercurrent in a Josephson junction.
We estimate infidelities below the surface code threshold $\sim 10^{-3}$~\cite{PhysRevLett.109.180502} without requiring  additional technologically demanding tuning of the individual Zeeman energies~\cite{Li2012, Wardrop2014, Buterakos2018, Cerfontaine20201} nor fine-tuning of parameters~\cite{Levy2002, Klinovaja2012, Mehl2014}. 
Our two-qubit gate design is fully compatible with current technology \cite{Jirovec2021}, could push ST qubits towards higher coherence standards, and can boost
the growing field of hybrid superconducting-semiconducting quantum systems.

This article is structured as follows. In Sec. \ref{sec:setup}, we describe the model setup in terms of a Fermi-Hubbard Hamiltonian. In Sec. \ref{sec:singlet-triplet-2-qubit-gate}, we calculate the ST two-qubit interaction and explain how leakage and crosstalk can be suppressed by the SOI  and by the phase difference across a Josephson junction. We conclude our findings in Sec. \ref{sec:conclusion}. In Appendixes \ref{sec:app_overview} to \ref{sec:leakage_plot_specifications} we discuss a more generalized model including nonuniform SOI, locally different Zeeman fields, and an explicit description of the Josephson junction.

\begin{figure*}[t]
	\includegraphics[width=0.98\textwidth]{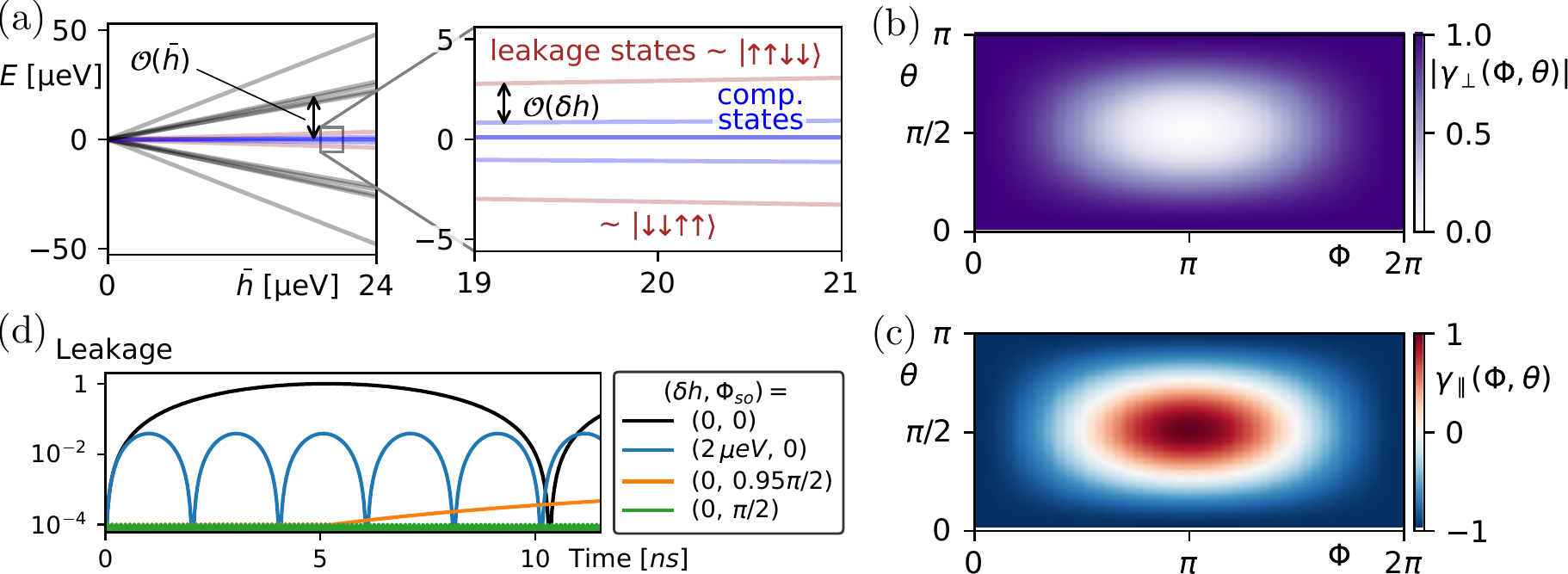}
	\caption{ %
 Leakage to noncomputational basis states.
 (a) Energy spectrum of our setup, described by $H_{\text{spin}}$  [Eq. \eqref{eq:Hspin}], as a function of the global Zeeman splitting $\bar{h}$ for $\mathcal{J}=0.4\,\mu$eV, $J_1=J_2=0$, and $\Phi_{\text{so}}=0$ (no SOI). States with total spin %
 $z-$component $S^z=1$ and $S^z=2$ (gray lines) are separated from the six $S^z=0$ states by the large energy $\mathcal{O}(\bar{h})$. In the inset, we highlight the two $S^z=0$ states (red lines) not belonging to the computational space. These states have energies $\mathcal{O}(\delta h)$, comparable to the computational states and cause large leakage in current devices. We use here  $(\delta h, \delta h_1, \delta h_2)/\bar{h}=(1/10,~1/20,~1/20)$. %
(b) The coupling $\mathcal{J}\gamma_{\perp}(\Phi)$ to the leading leakage states, and (c) the effective qubit-qubit interaction $\mathcal{J}\gamma_{\parallel}(\Phi)$ defined in Eq.~\eqref{eq:gamma_parallel_perp}. For large SOIs, the SOI rotation angle can become $\Phi_{\text{so}}=\pi
/2$ and the leakage coupling $\mathcal{J}\gamma_{\perp}(2\Phi_{\text{so}})$ vanishes when the Zeeman field $\vec{n}_B$ is perpendicular to the SOI vector $\vec{n}_\text{so}$ ($\theta=\pi/2$). At this operational sweet spot, the absolute qubit-qubit interaction ($|\mathcal{J}\gamma_{\parallel}(2\Phi_\text{so})|$) is also maximal, yielding the fastest possible two-qubit gates. 
(d) Leakage as a function of time $t$ in systems with large ($\Phi_{\text{so}}\approx \pi/2$) and without ($\Phi_{\text{so}}=0$) SOI. Without SOI, leakage is large and becomes negligible only at specific system-dependent times (black line) or when a large Zeeman energy difference between dot 2 and 3 is engineered (blue line). In our setup, leakage is orders of magnitude smaller (green line) at the sweet spot, and it  remains small also without fine-tuning the device (orange line). We use here the same parameters as (a) with  $\bar{h}=20\,\mu$eV; see also Appendix \ref{sec:leakage_plot_specifications}.
}\label{fig:423049850445989}
\end{figure*}

\section{Setup}\label{sec:setup}
We consider two ST qubits \cite{Levy2002, Petta2005,Jirovec2022, Jirovec2021, Jock2018, Fedele2021,Mutter2021} tunnel coupled via a superconducting lead; see Fig.~\ref{pic:13209801293}(a). 
An ST qubit comprises two exchange-coupled quantum dots, each containing %
a single spin, and is accurately modeled \cite{Mutter2021, PhysRevResearch.3.013081, Stepanenko2012} by the Fermi-Hubbard Hamiltonian 
\begin{align}
H&_{\text{DQD}}\label{eq:HDQDsp}\\
=&\sum_{\alpha\sigma\sigma'}\!\left(\!\epsilon_{\alpha}\delta_{\sigma\sigma'}\!+ \! \frac{1}{2} \left(\vec{h}_{\alpha}\cdot\vec{\sigma}\right)_{\sigma\sigma'}\!\right)\! d_{\alpha\sigma}^{\dagger} d_{\alpha \sigma'} +\! \mathcal{U} \sum_{\alpha}n_{\alpha\uparrow}n_{\alpha\downarrow}\nonumber\\\nonumber
&+t_0\sum_{\substack{ \sigma\sigma'\\ \alpha\in\{1,3\}}}\left(U_\text{so}^{\sigma\sigma'}(\Phi_\text{so})\,d_{\alpha+1,\sigma}^{\dagger}d_{\alpha\sigma'} +\text{H.c.}\right). 
\end{align}
Here,  $d_{\alpha\sigma}^{\dagger}$ creates a particle with spin $\sigma\in\{\uparrow,\downarrow\}$ on the dot $\alpha=\{1,2\}$ ($\alpha=\{3,4\}$) for the first (second) qubit with energy $\epsilon_{\alpha}<0$ and Kronecker delta $\delta_{\sigma\sigma'}$. The spin states are split by the Zeeman field $\vec{h}_\alpha=\mu_B\hat{g}_\alpha \vec{B}$, with $g$ tensor $\hat{g}_\alpha$, produced by an applied magnetic field $\vec{B}$. Double-occupation of each dot, specified by the occupation numbers $n_{\alpha\sigma}=d_{\alpha\sigma}^\dagger d_{\alpha\sigma}$, costs the on-site Coulomb energy $\mathcal{U}$.
Crucially,  $H_{\text{DQD}}$ includes tunneling events between the dots. These are parameterized by a real-valued tunneling amplitude $t_0>0$ and by a SOI-induced spin-flip operator $U_\text{so}(\Phi_\text{so})=\exp\left(i\Phi_\text{so} \vec{n}_\text{so}\cdot \pmb{\sigma}/2\right)$  that rotates the spins around the SOI vector $\vec{n}_\text{so}$ by the angle $\Phi_\text{so}\approx 2 L/l_{so}$~\cite{geyer2022two,PhysRevB.80.041301,yu2022strong}, with dot-dot distance $L$, SOI length $l_{so}$ and Pauli vector $\vec{\sigma}$. %
We emphasize that $\Phi_\text{so}$ is widely tunable by electrically controlling the position of the dots or the amplitude of the SOI \cite{Froning2021b}\footnote{In a general semiconducting material, the
spin-orbit interaction we consider is the combined effect of Rashba and Dresselhaus spin-orbit interaction.}; large values of $\Phi_\text{so}\sim \pi$, corresponding to complete spin flips, were recently measured in hole systems \cite{geyer2022two}. %

Superconductors are modeled by the mean-field BCS Hamiltonian
\begin{equation}\label{eq:H_Supercond_main}
 H_{\text{S}}=\sum_{\vec{k}\sigma}\epsilon_k c_{\vec{k}\sigma}^{\dagger} c_{\vec{k}\sigma} - \sum_{\vec{k}}\Delta c_{\vec{k}\uparrow}^{\dagger}c_{-\vec{k}\downarrow}^{\dagger}+\text{H.c.}\ , 
\end{equation}
 where $c_{\vec{k}\sigma}^{\dagger}$ creates an electron  with wave vector $\vec{k}$, spin $\sigma$, in a superconductor with superconducting gap $\Delta>0$ and normal-state energy $\epsilon_k$. In the following, %
  $\epsilon_k$ and $\epsilon_{\alpha}$ are measured with respect to the chemical potential.

 Each ST qubit is tunnel coupled to the superconductor as shown in Fig.~\ref{pic:13209801293}(a), described by the Hamiltonian
 \begin{align} \label{eq:H_T_main}
H_{\text{T}}&= t_{S} \sum_{\vec{k}\sigma\sigma'}U_\text{so}^{\sigma\sigma'}(\Phi_\text{so})\left(  c_{\vec{k}\sigma}^{\dagger} d_{2\sigma'}+  d_{3\sigma}^{\dagger}c_{\vec{k}\sigma'}\right)  +\text{H.c.}
\end{align}
In analogy to Eq.~\eqref{eq:HDQDsp}, we account for the SOI-induced spin flip by the rotation $U_\text{so}(\Phi_{\text{so}})$. 
We assume that all spin flips occur with the same angle $\Phi_\text{so}$ and direction $\vec{n}_\text{so}$. This corresponds to uniform SOI throughout the device and equal distances between QDs and between QDs and superconductors, a realistic scenario in experiments. %
In Appendixes \ref{sec:app_overview} to \ref{sec:arbitr_vec_h}, we analyze the general case with different spin rotation angles and axes, which will leave the main results unchanged, and where we extend our model to Josephson junctions  [Fig.~\ref{pic:13209801293}(b)].%

\section{Singlet-triplet two-qubit gates}\label{sec:singlet-triplet-2-qubit-gate}
\subsection{Superconductor-mediated exchange interactions}
The tunnel coupling to the superconductors affects the spin states confined in the quantum dots \cite{Choi2000,GonzalezRosado2021, Kornich2019, Kornich2020a, Spethmann2022}.
To the lowest order in the tunneling amplitude, the relevant mechanisms affecting the dots are elastic cotunneling \cite{Matsuo2022}, where one particle tunnels from one dot to the next through an electronic excitation of the superconductor, as well as local and crossed Andreev processes \cite{Deacon2015,kurtossy2022parallel}, where Cooper pairs are split and recombined in the same and different dots, respectively \cite{Scherubl2019}.
In the regime where $\Delta$ and $\mathcal{U}$ are large (see also Appendix \ref{sec:large_Coulomb_rep}), the elastic cotunneling  and local Andreev processes are suppressed, and we can focus on crossed Andreev processes only \cite{Scherubl2019}.

Crossed Andreev processes cause effective spin-spin correlations between distant dots,
\begin{equation}\label{eq:H_CA_maintext}
H_{\text{CA}}= \Gamma_\text{CA}\left(\begin{array}{c} -d_{3\downarrow}\\d_{3\uparrow}\end{array}\right)^T U_\text{so}(2\Phi_\text{so}) \left(
\begin{array}{c}
d_{2\uparrow}\\ d_{2\downarrow}\end{array}\right)+ \text{H.c.} %
\end{equation} 
The spin rotation $U_\text{so}(2\Phi_\text{so})$ combines two of the SOI rotations in $H_T$ [Eq.~\eqref{eq:H_T_main}]. The coupling strength is $\Gamma_\text{CA}=\pi t_S^2\rho_F$%
, where $\rho_F$ is the normal density of states per spin of the superconductors. %
We consider the width of the superconductors $w$ to be smaller than the superconducting coherence length $\xi$, in the micrometer range~\cite{Kittel2021, Mayer2019}. Beyond that, the resulting interaction decreases exponentially $\propto e^{-2w/\xi}$ \cite{Choi2000}. 

When all quantum dots are occupied by a single particle in their ground state, we derive the effective four-spin Hamiltonian for small $t_{0},\ \Gamma_{\text{CA}},\ |\vec{h}_{\alpha
}|\ll |\epsilon_{\alpha}|,\ \mathcal{U}$, \footnote{In the Appendixes we show that leakage reduction and vanishing crosstalk can be achieved even if $h_{\alpha}\sim \epsilon,\mathcal{U}$
} 
\begin{align}
\label{eq:Hspin}
H_{\text{spin}}=&\frac{1}{2}\sum_{\alpha}  \vec{h}_{\alpha}\cdot\pmb{\sigma}^{\alpha} +\frac{\mathcal{J}}{4} \pmb{\sigma}^{2}\cdot \hat{R}_\text{so}(2\Phi_\text{so}) \pmb{\sigma}^{3}\nonumber\\
&+\frac{ J_1}{4}\pmb{\sigma}^{1}\cdot \hat{R}_\text{so}(\Phi_\text{so}) \pmb{\sigma}^{2}
+\frac{ J_2}{4} \pmb{\sigma}^{3}\cdot \hat{R}_\text{so}(\Phi_\text{so}) \pmb{\sigma}^{4} ,
\end{align}
 with spin operators  $\vec{\sigma}^{\alpha}$ of dot $\alpha$, energy detunings $\tilde{\epsilon}_1=\epsilon_{1}-\epsilon_{2}$ and $\tilde{\epsilon}_2=\epsilon_3-\epsilon_4$, and coupling constants
\begin{equation}
J_i= \frac{4t_0^2 \mathcal{U}}{\mathcal{U}^2-\tilde{\epsilon}_i^2} \  \text{and} \  \mathcal{J}= \frac{-4\Gamma_\text{CA}^2 \mathcal{U}}{(\epsilon_2+\epsilon_3)(2\mathcal{U}+\epsilon_2+\epsilon_3)}    .\label{eq:couplingconstants}
\end{equation} 
Importantly, the exchange interactions are anisotropic and are given by the rotation matrices $\hat{R}_{\text{so}}(\Phi)$, which describe right-handed rotations around the vector $\vec{n}_\text{so}$ of an angle $\Phi$~\cite{geyer2022two}. %
 The energy spectrum of $H_\text{spin}$, highlighting the relevant computational states, is shown in Fig.~\ref{fig:423049850445989}(a), and depends on the global, averaged Zeeman splitting $\bar{h}$ as well as the Zeeman energy differences $\delta h_1=h_1-h_2$, $\delta h=h_2-h_3$, and $\delta h_2=h_3-h_4$ (assuming parallel $\vec{h}_\alpha=h_\alpha \vec{n}_B$).

We anticipate that in the setup sketched in Fig.~\ref{pic:13209801293}(b), where a single superconductor is substituted by a Josephson junction, the effective exchange $\mathcal{J}$ becomes externally controllable by the superconducting phase difference $\varphi$, enabling on-demand switching  on and off of these interactions; see Appendix \ref{appendix:sec:calculate_H_eff}.
We also emphasize that this effective model [Eq.~\eqref{eq:Hspin}] is equivalent to a chain of four QDs that are directly coupled by exchange interactions, as shown in Fig.~\ref{pic:13209801293}(c). Consequently, our approach to reduce leakage and to achieve high-fidelity two-qubit gates is valid also in these systems. 

\subsection{Singlet-triplet qubit coupling}
We now derive the effective coupling between our two ST qubits. To simplify the discussion, here we assume that the Zeeman fields of each spin $\alpha$ are aligned along the direction $\vec{n}_B$ but can change in magnitude, i.e., $\vec{h}_\alpha=h_\alpha \vec{n}_B$; the general case is discussed in Appendix \ref{sec:arbitr_vec_h} and only the rotation matrices $\hat{R}_\text{so}$ are renormalized. %

When the global Zeeman field is much larger than the exchange couplings, we can project $H_\text{spin}$ onto the computational subspace of the ST qubits
yielding the two-qubit Hamiltonian
 \begin{equation}
H_{\text{ST}}=\frac{1}{2}\vec{\mathcal{B}}_1\cdot \vec{\tau}^1 + \frac{1}{2}\vec{\mathcal{B}}_2 \cdot \vec{\tau}^2 + \frac{\mathcal{J}\gamma_{\parallel}(2\Phi_\text{so})}{4} \tau_z^1 \tau_z^2 \ ,\label{eq:098345034542}
 \end{equation}
 where $\vec{\tau}^1$  ($\vec{\tau}^2$) is the Pauli vector acting on the computational space of the first (second) ST qubit, spanned by the states \{$|\!\!\uparrow_1\downarrow_2\rangle, |\!\!\downarrow_1\uparrow_2\rangle$\} \big(\{$|\!\!\uparrow_3\downarrow_4\rangle$, $|\!\!\downarrow_3\uparrow_4\rangle$\}\big), with $|\!\uparrow_{\alpha}\rangle$ pointing along $\vec{n}_B$.
 The single-qubit terms $\vec{\mathcal{B}}_i=\left( J_i \text{Re}\left[\gamma_{\perp}(\Phi_\text{so})\right] ,   J_i \text{Im}\left[\gamma_{\perp}(\Phi_\text{so})\right] ,   \delta h_i\right)$ enable single qubit rotations.
 The superconductor mediates effective Ising qubit-qubit interactions $\propto \mathcal{J}\gamma_{\parallel}(2\Phi_\text{so})$ via crossed Andreev reflection. Single- and two-qubit gates are enabled by controlling $J_i$ and $\mathcal{J}$, respectively.
 Importantly, the SOI induces an anisotropic factor
 described by the dimensionless functions
 \begin{align}\label{eq:gamma_parallel_perp}
     \gamma_\parallel(\Phi) &= 2\sin^2(\theta)\sin^2(\Phi/2)-1\ ,\nonumber\\
     \gamma_\perp(\Phi) &=  \left[\cos\left(\Phi/2\right)+i\cos(\theta)\sin\left(\Phi/2\right)\right]^2.%
 \end{align}
These functions are shown in Figs.~\ref{fig:423049850445989}(b) and \ref{fig:423049850445989}(c), respectively. We emphasize that they are fully controllable by the SOI angle $\Phi_\text{so}$ and by the direction $\theta$ of the Zeeman vector $\vec{n}_B$ relative to the SOI vector $\vec{n}_\text{so}$ (Fig.~\ref{pic:13209801293}).

\begin{figure*}
	\includegraphics[width=0.98\textwidth]{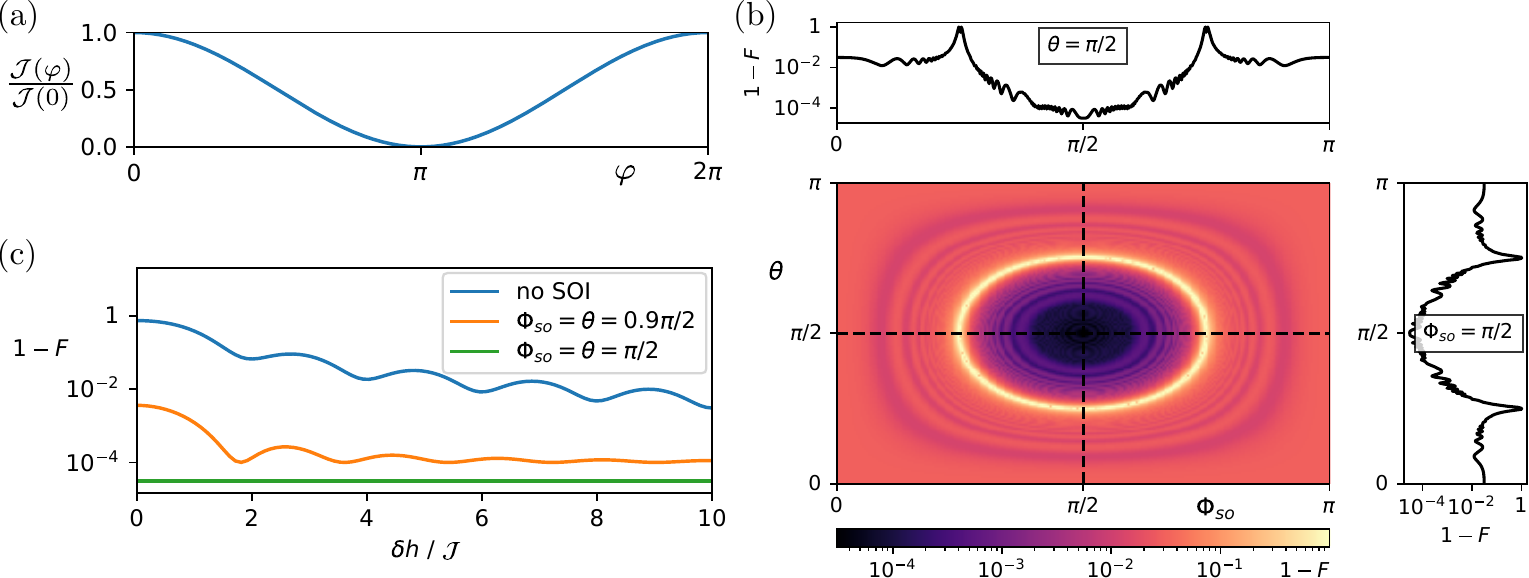}
	\caption{ Two-qubit gates.
 (a)  Control of qubit-qubit interactions $\mathcal{J}$ by tuning the superconducting phase difference $\varphi$; see Fig.~\ref{pic:13209801293}(b). The interaction can be precisely switched on and off, removing residual interactions and crosstalk that hinder scalability.
In current devices $\mathcal{J}\approx 0.4$~$\mu$eV, so short gate times of $T_g\sim 5$~ns are within reach.
(b,c) Two-qubit gate infidelity $1-F$. (b) The large SOI enables high-fidelity gates for a wide range of parameters close to the sweet spot at $\Phi_{\text{so}}=\theta=\pi/2$. $\delta h/\mathcal{J}=5$. (c)
Comparing the performance of different approaches, we observe that our setup consistently outperforms other devices, yielding fidelities orders of magnitude larger even without precisely fine-tuning the system to the sweet spot.
We use $\bar{h}/\mathcal{J}=50$, $J_{1}=J_{2}=0$, and $(\delta h_1,~\delta h_2)/\delta h=(1/2,~ 1/2)$.}\label{fig:3498578236987}
\end{figure*}

\subsection{Leakage suppression by spin-orbit interaction}
The projection onto the ST qubit subspace leading to $H_\text{ST}$ is valid when the noncomputational states are decoupled from the computational space. In a single ST qubit and at typical values of $\vec{B}$, the noncomputational subspace is well separated by the large total Zeeman energy $h_{\alpha}\sim5$GHz, orders of magnitude larger than the characteristic energy of $H_\text{ST}$, in the $10$ to $100$ MHz range. %
However, in two ST qubits, there are two noncomputational states in the computational energy window, resulting in large leakage. These states  ($|\!\uparrow_1\uparrow_2\downarrow_3\downarrow_4\rangle$ and $|\!\downarrow_1\downarrow_2\uparrow_3\uparrow_4~\rangle$) are not affected by the large total Zeeman energy because they have zero total spin, $S^z=0$; see Fig.~\ref{fig:423049850445989}(a). 

This critical and fundamental flaw of ST architectures was addressed previously by fine-tuning isotropic exchange interaction and magnetic field such that the leakage vanishes at the target evolution time~\cite{Levy2002, Klinovaja2012, Mehl2014}; see black and blue curves in Fig.~\ref{fig:423049850445989}(d). 
Alternatively, leakage can be partially reduced by engineering a large Zeeman energy difference $\delta h$~\cite{Li2012, Wardrop2014, Buterakos2018, Cerfontaine20201}.
These approaches, however, require extremely precise control over  $g$ factors and over exchange, %
which is challenging to achieve in current experiments.
Capacitive and resonator-mediated couplings \cite{Shulman2012,Nichol2017,PhysRevB.100.035416,Bottcher2022} \cite{Leijnse2013} yield only weak interaction strengths, are more susceptible to charge noise, and result in slow gates.

In striking contrast, in our system the strong SOI offers 
a compelling way  to  remove leakage, fully compatible with the state of the art~\cite{geyer2022two}.
While the qubit-qubit interactions are determined by $\mathcal{J}\gamma_\parallel(2\Phi_{\text{so}})$,  leakage is determined by the matrix elements that couple the computational states $|\!\uparrow_1\downarrow_2\uparrow_3\downarrow_4\rangle$ and $|\!\downarrow_1\uparrow_2\downarrow_3\uparrow_4\rangle$ to the noncomputational states $|\!\uparrow_1\uparrow_2\downarrow_3\downarrow_4~\rangle$ and $|\!\downarrow_1\downarrow_2\uparrow_3\uparrow_4\rangle$, and are given by $ \mathcal{J}\gamma_\perp(2\Phi_{\text{so}})$ (see Appendix \ref{sec:leakage_coupling}).
By operating the systems at $\Phi_{\text{so}}=\pi/2$ and at $\vec{n}_B\perp\vec{n}_\text{so}$ ($\theta=\pi/2$), one maximizes the two-qubit interaction [$\mathcal{J}\gamma_{\parallel}(\pi)=\mathcal{J}$] and suppresses leakage [$\mathcal{J}\gamma_{\perp}(\pi)=0$].
At this operational sweet spot, particles from dots 2 and 3 make a rotation of $\pi/2$ when tunneling to the superconductor, but in opposite directions. Since Cooper pairs can only be created from particles in the superconductor with opposite spin, $|\!\!\downarrow_2\uparrow_3\rangle$ and $|\!\!\uparrow_2\downarrow_3\rangle$ will not be affected by crossed Andreev processes. In contrast, $|\!\!\downarrow_2\downarrow_3\rangle$ and $|\!\!\uparrow_2\uparrow_3\rangle$ will couple to Cooper pairs and experience a shift in energy, resulting in the Ising qubit-qubit interaction required for two-qubit gates.
Although the anisotropic exchange also causes leakage to states with $S^z\neq0$ (e.g. $|\!\!\uparrow_1\uparrow_2\uparrow_3\uparrow_4\rangle$), this contribution is small because those states are at high energies, determined by the large \textit{global} Zeeman field; see Fig.~\ref{fig:423049850445989}(a). In Appendixes \ref{sec:large_Zeeman_and_nonuniform_SC} and \ref{sec:correct_leak} we calculate higher-order corrections to the qubit-qubit interaction and the leakage coupling.

As shown in Fig.~\ref{fig:423049850445989}(d), our SOI-induced leakage suppression can significantly outperform  current alternative approaches, removing also technologically demanding constraints on the engineering of the devices.
We also emphasize that while leakage is minimal at $\Phi_{\text{so}}=\pi/2$ and $\vec{n}_B\perp\vec{n}_\text{so}$, conditions achieved in recent experiments~\cite{geyer2022two}, our system does not require precise fine-tuning and leakage is significantly lower than the state of the art for a wide range of $\Phi_{\text{so}}$.

Finally, we stress that leakage reduction is compatible with high-fidelity single-qubit gates. While leakage coupling proportional to $\gamma_{\perp}(2\Phi_{\text{so}})=0$ vanishes at the sweet spot $\Phi_{\text{so}}=\pi/2$, the term $\gamma_{\perp}(\Phi_{\text{so}})$, which enters the single qubit term $\mathcal{B}_i$ in Eq.~\eqref{eq:098345034542}, remains nonzero. In fact, the spin-orbit rotation angles that determine the single-qubit gates and those that determine the two-qubit leakage term can be tuned independently from each other, as shown in detail in Appendixes \ref{sec:general_model} and \ref{appendix:sec:calculate_H_eff}.

\subsection{Controlling exchange by Josephson junctions}
The other critical obstacle to scaling up current ST architectures is the residual exchange interaction between two qubits, which yields a dangerous always-on coupling between ST qubits.
Our hybrid device sketched in Fig.~\ref{pic:13209801293}(b) removes addresses this issue by taking full advantage of the phase tunability of Josephson junctions. In this case, the effective exchange interaction $\mathcal{J}$ becomes dependent on the superconducting phase difference $\varphi$, yielding $\mathcal{J}\to \mathcal{J}(\varphi)=4\mathcal{J}\cos^2\left(\frac{\varphi}{2}\right)$, %
see Appendix \ref{appendix:sec:calculate_H_eff} for a detailed derivation. %
This interaction is thus maximal when $\varphi=0$, and vanishes at $\varphi=\pi$, as  %
 shown in Fig.~\ref{fig:3498578236987}(a). %
 Because $\varphi$ is accurately controllable in experiments, our setup offers a long-sought way to on-demand switch interactions on and off, removing the main source of crosstalk in future large-scale  ST qubit architectures.

\subsection{High-fidelity two-qubit gates} 
The SOI-induced anisotropy and the phase-tunability of the effective exchange interactions in our setup 
enable fast and high-fidelity two-qubit gates; see Figs.~\ref{fig:3498578236987}(b) and \ref{fig:3498578236987}(c).
By assuming $\epsilon_{\alpha}\approx -50$~$\mu$eV, $\Gamma_\text{CA}\approx 4.5$~$\mu$eV and $\mathcal{U},~\Delta\gg|\epsilon_{\alpha}|$, we estimate that $\mathcal{J}$ in Eq.~\eqref{eq:couplingconstants} can reach realistic values up to $\mathcal{J}\approx 0.4$~$\mu$eV. 
Fast controlled-Z (cZ) entangling gates are then enabled by turning the interactions on for a time $T_g=\hbar\pi/|\mathcal{J}\gamma_{\parallel}(2\Phi_{\text{so}})|$, which may take around $T_g\approx 5$~ns only. %
We estimate  the fidelity  by
\begin{align}
F=\left|\frac{1}{4}\text{tr}(U_{\text{cZ}}^{\dagger}U_{\text{spin}})\right|^2 ,
\end{align}
where $U_{\text{cZ}}=\exp(-i\pi\tau_z^1\tau_z^2/4)$ is the ideal cZ gate (up to single-qubit operations \cite{Loss1998}) and $U_{\text{spin}}=P_\text{comp} \exp(-i H_\text{spin} T_g/\hbar) P_\text{comp}$ is the time evolution  generated by $H_\text{spin}$ in Eq.~\eqref{eq:Hspin} projected onto the computational subspace by the projection operator $P_\text{comp}$.   Our approach accurately captures leakage because $H_\text{spin}$ includes all $2^4$ spin states.

As shown in Fig.~\ref{fig:3498578236987}(c), our ST qubits substantially outperform current state of the art, reaching two orders of magnitude smaller values of infidelities, below $10^{-4}$ at the optimal parameter spot. This value is limited by our conservative choice of the global Zeeman energy $\bar{h}=50\mathcal{J}\sim 20\,\mu$eV, in contrast to current implementations where the fidelity is limited by $\delta h\ll \bar{h}$.
The infidelity remains below $10^{-3}$ for small Zeeman energy differences $\delta h$ at values of $\Phi_{\text{so}}$ and $\theta$ deviating up to $10\%$ from the sweet spot, demonstrating that our approach does not require precise fine-tuning of the device, and that fidelities larger by more than two orders of magnitude are within reach in current experiments.

\section{Conclusion}\label{sec:conclusion}
In conclusion, hybrid ST qubit architectures comprising semiconducting quantum dots with large SOI and superconductors can substantially outperform current devices. In particular, the superconductor mediates correlations of distant qubits via crossed Andreev processes. These processes are externally controllable by the phase difference in Josephson junctions, 
removing  dangerous crosstalk caused by always-on residual qubit-qubit interactions.
We also show that large SOIs induce tunable anisotropies in these interactions  that strongly suppress leakage. When combined, these effects result in fast and high-fidelity two-qubit gates, orders of magnitude more efficient than the state of the art, and could provide a significant step forward toward implementing large-scale ST qubit quantum processors.

\section{Acknowledgements}
 This project has received funding from the European Union's Horizon 2020 research and innovation program under Grant Agreement No 862046 and under Grant Agreement No 757725 (the ERC Starting Grant). This work was supported by the Swiss National Science Foundation, NCCR QSIT, and NCCR SPIN (Grant No. 51NF40-180604). Finally, this work was financially supported by the JSPS Kakenhi Grant No. 19H05610.

\setcounter{section}{0}
\setcounter{equation}{0}
\appendix

\section{Overview}\label{sec:app_overview}

In the appendixes presented hereafter we give more details on the calculations presented in the main text. In Appendix \ref{sec:general_model}, we describe a more generalized Hamiltonian that models our ST qubits including a Josephson junction and with less uniform SOI parameters. In Appendix \ref{appendix:sec:calculate_H_eff}, we describe the crossed Andreev processes that result in an effective spin Hamiltonian, and eventually in a two-qubit interaction of the ST qubits, tunable by the superconducting phase difference and the SOI parameters. In Appendix \ref{sec:arbitr_vec_h}, we explain why local differences in the Zeeman field only cause a renormalization of the spin-orbit rotations, and in Appendix \ref{sec:large_Coulomb_rep} we show that our results stay valid in the regime of large Coulomb interaction. We derive the coupling to the leakage states in Appendix \ref{sec:leakage_coupling}.  In Appendix \ref{sec:large_Zeeman_and_nonuniform_SC}, we show that crosstalk and leakage are suppressed also for larger Zeeman energies and that this suppression is robust even when the quantum dots do not couple equally to both superconductors. In Appendix \ref{sec:correct_leak}, we calculate how higher-order corrections affect the leakage and the two-qubit interaction. In Appendix \ref{sec:leakage_plot_specifications}, we give more details of how we calculate the leakage plot Fig.~\ref{fig:423049850445989}(d).

\section{Generalized model of the ST qubits}\label{sec:general_model}
We start with a general model of our setup, described by $\mathcal{H}=\mathcal{H}_{\text{DQD}}+\mathcal{H}_{\text{S}}+\mathcal{H}_{\text{T}}$, and include a Josephson junction with two superconducting leads instead of a single superconducting lead. We also consider the effects of less uniform spin-orbit parameters and dot-dot distances throughout the device. 
The part related to the double quantum dot $\mathcal{H}_{\text{DQD}}$ is a generalized version of $H_\text{DQD}$ [Eq.~\eqref{eq:HDQDsp}] \cite{PhysRevResearch.3.013081}, 
\begin{align}
&\mathcal{H}_{\text{DQD}}=\sum_{\alpha\sigma} \left(\epsilon_{\alpha}+ \sigma\frac{h_{\alpha}}{2} \right) d_{\alpha\sigma}^{\dagger} d_{\alpha \sigma} +\! \mathcal{U} \sum_{\alpha}n_{\alpha\uparrow}n_{\alpha\downarrow}\nonumber\\
&\hspace{10pt}+\sum_{\sigma\sigma'} \left(t_{1}U^{\sigma\sigma'}_{1}\,d_{2\sigma}^{\dagger}d_{1\sigma'} + t_{2}U^{\sigma\sigma'}_{2}\,d_{4\sigma}^{\dagger}d_{3\sigma'} +\text{H.c.}\right).\label{eq:HDQDsp2}
\end{align}
The difference is that here for each double quantum dot $i\in{1,2}$ we have different real-valued tunneling amplitudes $t_i>0$ and different SOI-induced spin-flip operators $U_{i}=\exp\left(i\Phi_i \vec{n}_i\cdot \pmb{\sigma}/2\right)$  that rotate the spins around the SOI vector $\vec{n}_i$ by the angle $\Phi_i$. The angle $\Phi_i$ is related to the distance between the corresponding dots, thus $\Phi_1\neq\Phi_2$ accounts for nonequal dot-dot distances. We choose the spin basis of each dot such that the Zeeman field $h_{\alpha}$ is parallel to the $z$ direction ($\sigma\in\{\uparrow=1,\downarrow=-1\}$ accounts for the proper sign). In Appendix \ref{sec:arbitr_vec_h}, we will discuss the implications of this choice in systems with arbitrary, dot-dependent $g$ tensors.
The superconductors are modeled by the mean-field BCS Hamiltonian
\begin{equation}\label{eq:H_S_SM}
 \mathcal{H}_{\text{S}}=\sum_{j\vec{k}\sigma}\epsilon_k c_{j\vec{k}\sigma}^{\dagger} c_{j\vec{k}\sigma} - \sum_{j\vec{k}}\Delta e^{-i\varphi_j} c_{j\vec{k}\uparrow}^{\dagger}c_{j-\vec{k}\downarrow}^{\dagger}+\text{H.c.}\ , 
\end{equation}
 where compared to $H_\text{S}$ [Eq.~\eqref{eq:H_Supercond_main}], we account for two superconductors, an upper and a lower one, which differ by their superconducting phase $\varphi_j$ and are indexed with $j\in\{u,l\}$. 
Each singlet-triplet  qubit is tunnel-coupled to the superconductors, described by 
 \begin{align}\label{eq:H_T_SM}
\mathcal{H}_{\text{T}}&=  \sum_{j\vec{k}\sigma\sigma'}\left(t_{j2} U^{\sigma\sigma'}_{j2}c_{j\vec{k}\sigma}^{\dagger} d_{2\sigma'}+  t_{j3} U^{\sigma\sigma'}_{j3} d_{3\sigma}^{\dagger}c_{j\vec{k}\sigma'}  +\text{H.c.}\right),
\end{align}
where, in analogy to Eq.~\eqref{eq:HDQDsp2}, we have real-valued tunnel amplitudes $t_{j\alpha}>0$ and we account for the SOI-induced spin-flip by the rotations $U_{j\alpha}=\exp(i\Phi_{j\alpha}\vec{n}_{j\alpha}\cdot\vec{\sigma}/2)$, that can in principle differ from each other. 
Note that the particles described by $d_{\alpha\sigma}^{\dagger}$ and $c_{j\vec{k}\sigma}^{\dagger}$ in our Hamiltonian, can be both an electron or a hole.

\section{Calculation of the effective singlet-triplet Hamiltonian}\label{appendix:sec:calculate_H_eff}

We  derive the effective Hamiltonian $\mathcal{H}_\text{ST}$ in three steps. In each step we apply Schrieffer-Wolff perturbation theory \cite{Winkler2003,Bravyi2011} to decrease the number of degrees of freedom. In the first step, we calculate how crossed Andreev processes described by $\mathcal{H}_\text{CA}$ [Eq.~\eqref{eq:H_CA_maintext}] affect the dot states assuming  a large superconducting gap $\Delta$. 
In the next step, we  obtain a spin Hamiltonian $\mathcal{H}_\text{spin}$ [Eq.~\eqref{eq:Hspin}] assuming a symmetry between the tunneling to the upper and lower superconductor given by $U_{u3}U_{u2}=U_{l3}U_{l2}=U$ and $t_{u2}t_{u3}=t_{l2}t_{l3}=t^2$, and assuming that the dots are occupied by one particle only. In the third step, we obtain the effective ST Hamiltonian in $\mathcal{H}_\text{ST}$ [Eq.~\eqref{eq:098345034542}]. The parameters are assumed to fulfill $t,\, h_{\alpha}\ll |\epsilon_{\alpha}|,\ \mathcal{U}\ll \Delta$. The setup with only one superconductor can be obtained by decoupling the second superconductor by choosing $t_{l\alpha}=0$. In Appendix \ref{sec:large_Coulomb_rep}, we argue why the condition $\mathcal{U}\ll\Delta$ can be revoked without affecting the results. %
In Appendix \ref{sec:large_Zeeman_and_nonuniform_SC}, we derive that leakage and crosstalk are suppressed even when the coupling to the two superconductors is nonuniform (thus, when we have deviations from $U_{u3}U_{u2}=U_{l3}U_{l2}$ and $t_{u2}t_{u3}=t_{l2}t_{l3}$) and when the Zeeman energy is of the order of the other dot energies (thus, covers $h_{\alpha}\not\ll|\epsilon_{\alpha}|,\mathcal{U}$). This calculation can be followed without having read Apps. \ref{sec:spin_Hamilt} - \ref{sec:leakage_coupling} and is more straight-forward but only describes the system near our optimal point, which corresponds to $\theta=\pi/2$ and $\Phi_\text{so}=\pi/2$.

\subsection{Quantum dot Hamiltonian}\label{sec:QD_Hamilt}

In this section, we show how the superconductor mediates an interaction between particles on the dots $2$ and $3$ based on crossed Andreev processes $\mathcal{H}_\text{CA}$ [Eq.~\eqref{eq:H_CA_maintext}]. In this part, we will focus on dots $2$ and $3$ and closely follow Ref.~\cite{Scherubl2019}.
Using second-order perturbation theory with respect to the tunneling $\mathcal{H}_T$, we project onto a subspace where the superconductor is in its ground state. The contributions from the upper superconducting lead and the lower superconducting lead are independent of each other and can be calculated separately.
We first study the contributions from the upper superconducting lead and can then extend the results to the lower superconductor.
We start with a basis transformation in the spin space of the superconductor and of dot 2 that we denote with a  tilde, %
\begin{align}
\begin{pmatrix}\Tilde{c}_{u\vec{k}\uparrow}\\\Tilde{c}_{u\vec{k}\downarrow}\end{pmatrix} =U_{u3} \begin{pmatrix}c_{u\vec{k}\uparrow}\\c_{u\vec{k}\downarrow}\end{pmatrix},~
\begin{pmatrix}\Tilde{d}_{2\uparrow}\\\Tilde{d}_{2\downarrow}\end{pmatrix} =U_{u} \begin{pmatrix}d_{2\uparrow}\\d_{2\downarrow}\end{pmatrix}\label{eq:basis_trafo_d2}\ ,
\end{align}
with $U_{u}=U_{u3}U_{u2}$. While the superconductor and the on-site Coulomb interaction are invariant with respect to this basis transformation, the Zeeman splitting is rotated and now described by $\sum_{\sigma\sigma'}h_2(U_{u} \sigma^z U_u^{\dagger})_{\sigma\sigma'}\tilde{d}_{2\sigma}^{\dagger}\tilde{d}_{2\sigma'}/2$. Importantly, the part of $\mathcal{H}_T$ that couples to the upper superconductor, $\mathcal{H}_T^u$, in this locally rotated frame only consists of spin-conserving tunneling:
\begin{align}
\mathcal{H}_\text{T}^u=\sum_{\vec{k}\sigma}\big[t_{u2}\tilde{c}_{u\vec{k}\sigma}^{\dagger}\tilde{d}_{2\sigma}+ t_{u3}d_{3\sigma}^{\dagger} \tilde{c}_{u\vec{k}\sigma} \big] + \text{H.c.}
\end{align}
With a standard Bogoliubov transformation for the superconductor $\tilde{c}_{u\vec{k}\sigma}=u_k\gamma_{u\vec{k}\sigma}+\sigma v_{uk} \gamma_{u-\vec{k}-\sigma}^{\dagger}$ with $u_k=[(1+\epsilon_k/E_k)/2]^{1/2}$, $v_{uk}=\exp(-i\varphi_u)[(1-\epsilon_k/E_k)/2]^{1/2}$, where $E_k=\sqrt{\Delta^2+\epsilon_k^2}$, we find
\begin{align}
\mathcal{H}_\text{T}^u=\sum_{\vec{k}\sigma}\big[&t_{u2}(u_k\gamma_{u\vec{k}\sigma}^{\dagger}+\sigma v_{uk}^*\gamma_{-\vec{k}{-\sigma}})\tilde{d}_{2\sigma} \\\nonumber
&+ t_{u3}d_{3\sigma}^{\dagger}(u_k\gamma_{u\vec{k}\sigma} + \sigma v_{uk}\gamma_{u-\vec{k}{-\sigma}}^{\dagger})\big] + \text{H.c.}
\end{align}
Following Ref.~\cite{Scherubl2019}, we now derive an effective Hamiltonian describing the dot states using second-order perturbation theory in $\mathcal{H}_\text{T}$ and in the tunneling between the double quantum dots \footnote{we also include non-diagonal elements of the Zeeman field as a perturbation}. We assume $|\epsilon_{\alpha}|,\ \mathcal{U},\ h_{\alpha},\ t_{j\alpha }\ll\Delta$. 
For a Hamiltonian $H=H_0+H'$ with known eigenstates $|n\rangle$ and eigenavalues $E_n$ of $H_0$ and a perturbation $H'$, the effective Hamiltonian $H_\text{eff}$ acting on a  quasidegenerate subspace $\mathcal{A}$ well separated from the other states is given  to second order by \cite{Winkler2003}
\begin{align}\label{eq:092834753048}
\langle m|H_\text{eff}|m'\rangle=&\langle m| H_0+H'|m'\rangle\\ \nonumber
&+ \textstyle\sum\limits_l \frac{\langle m|H'|l\rangle\,\langle l|H'|m'\rangle}{2}\left(\frac{1}{E_m-E_l}+\frac{1}{E_{m'}-E_l}\right)\ ,
\end{align}
with $|m\rangle,|m'\rangle\in\mathcal{A}$, $|l\rangle\in\mathcal{A}^{\perp}$, where $\mathcal{A}^{\perp}$ is the space orthogonal to $\mathcal{A}$. We now choose $\mathcal{A}$ as the subspace where the superconductor is in its ground state $|G\rangle$ but we still allow for arbitrary dot states. The zeroth and first order of the perturbation theory is just the double dot Hamiltonian $\mathcal{H}_\text{DQD}$.
The second-order contribution comprises elastic cotunneling $\mathcal{H}_\text{EC}$, local Andreev processes $\mathcal{H}_\text{LA}$, and crossed Andreev processes $\mathcal{H}_\text{CA}$ \cite{Scherubl2019}. By evaluating these contributions, we find the effective  Hamiltonian
\begin{align}
 \mathcal{H}_\text{dot}=\mathcal{H}_\text{DQD}+\mathcal{H}_\text{CA}+\mathcal{H}_\text{LA}+\mathcal{H}_\text{EC}.
\end{align}
Local Andreev processes are those that create or destroy two electrons with opposite spins on the same dot.
Since we will eventually focus on those states where a single particle is on each dot and which are therefore not affected by $\mathcal{H}_\text{LA}$, this contribution will not play a role in our final result.
Elastic cotunneling $\mathcal{H}_\text{EC}$ transfers a particle from one dot through the superconductor to the other dot.
Since these contributions are suppressed when the energies ($\epsilon_{\alpha},\mathcal{U}$) are small compared to the superconducting gap $\Delta$ (see also Ref.~\cite{Scherubl2019}), we will neglect this term further and will focus on the crossed Andreev processes $\mathcal{H}_\text{CA}$.

 Crossed Andreev processes are those where a Cooper pair in the superconductor splits up and one particle tunnels to dot 2, the other one with opposite spin to dot 3. 
Therefore, $\mathcal{H}_\text{CA}$ consists of the matrix elements $\langle\tilde{\uparrow}_2 \downarrow_3G|\mathcal{H}_\text{CA}|0G\rangle$ and $\langle\tilde{\downarrow}_2\uparrow_3G|\mathcal{H}_\text{CA}|0G\rangle$ and their Hermitian conjugate. Here, $|\sigma_{\alpha}\rangle=d_{\sigma\alpha}^{\dagger}|0\rangle$ describes the state of a particle with spin $\sigma$  on dot $\alpha$ and $|\tilde{\sigma}_{\alpha}\rangle=\tilde{d}_{\sigma\alpha}^{\dagger}|0\rangle$ is the state in the rotated spin basis.
We evaluate the contribution coming from the upper superconductor: 
\begin{align}\label{eq:09834523024309}
\textstyle\langle\tilde{\uparrow}_2 &\downarrow_3G|\mathcal{H}_\text{CA}^u|0G\rangle
\\
&=\textstyle\frac{t_{u2}t_{u3}}{2}\sum\limits_{\vec{k}}u_kv_k \Big(\frac{1}{\epsilon_3-h_3/2-E_k}+ \frac{1}{-\epsilon_2-\Tilde{h}_2/2-E_k} \nonumber\\\nonumber
&\qquad\qquad\quad\qquad+ \textstyle\frac{1}{\epsilon_2+\Tilde{h}_2/2-E_k}+ \frac{1}{-\epsilon_3+h_3/2-E_k}\Big)\\\nonumber
&\textstyle\overset{\epsilon_{\alpha},h_{\alpha}\ll\Delta}{\approx}-2t_{u2}t_{u3}\sum\limits_{\vec{k}}\frac{u_k v_{uk}}{E_k}\\\nonumber
&=-t_{2u}t_{3u}e^{-i\varphi_u}\int \text{d}\epsilon\ \rho(\epsilon)  \frac{\Delta}{\Delta^2+\epsilon^2}\\\nonumber
&\approx -\pi t_{u2}t_{u3}\rho_Fe^{-i\varphi_u}\equiv-\Gamma_{\text{CA}, u}\nonumber\ ,
\end{align}
with $\Tilde{h}_2=h_2(U_{u} \sigma^z U_u^{\dagger})_{\uparrow\uparrow}$, assuming a constant normal density of states per spin $\rho_F$ and defining $\Gamma_{\text{CA},u}$ in the last line \footnote{
The result is independent of the occupation of the remaining spin-down level of dot $2$ or spin-up level of dot $3$%
, up to possible factors of $(-1)$ from fermionic commutators. For example, $\langle\tilde{\uparrow}_2\tilde{\downarrow}_{2}\downarrow_3 G|\mathcal{H}_\text{CA}^u|\tilde{\downarrow}_2 G\rangle=\Gamma_{\text{CA},u}$. Fermionic signs are chosen such that the state $|\uparrow_2\downarrow_2\uparrow_3\downarrow_3\rangle$ gets a positive sign when particles fill up from the right.}.
In analogy to Eq.~\eqref{eq:09834523024309}, we find $\langle\Tilde{\downarrow}_2 \uparrow_3G|\mathcal{H}_\text{CA}^u|0G\rangle=\Gamma_{\text{CA},u}$, resulting in
\begin{align}
    \mathcal{H}_\text{CA}^u=\Gamma_\text{CA,u}(\tilde{d}_{2\downarrow}^{\dagger}d_{3\uparrow}^{\dagger}-\tilde{d}_{2\uparrow}^{\dagger}d_{3\downarrow}^{\dagger})+ \text{H.c.}
\end{align}
When we rotate back to our original basis [Eq.~\eqref{eq:basis_trafo_d2}] and extend our calculation to the lower superconductor, we get (with $\Gamma_\text{CA,l}=\pi t_{l2}t_{l3}\rho_Fe^{-i\varphi_l}$ and $U_{l}=U_{l3}U_{l2}$)
\begin{align}\label{eq:H_CA_general}
\mathcal{H}_{\text{CA}}=\begin{psmallmatrix}d_{2\uparrow}^{\dagger}\\d_{2\downarrow}^{\dagger}\end{psmallmatrix}^T\big(\Gamma_{\text{CA},u}U_{u}^{\dagger}\!+\!\Gamma_{\text{CA},l}U_{l}^{\dagger}\big)\begin{psmallmatrix} -d_{3\downarrow}^{\dagger}\\d_{3\uparrow}^{\dagger}\end{psmallmatrix} + \text{H.c.}
\end{align}
In what follows we will focus on the simplest realistic case where $t_{l2}t_{l3}=t_{u2}t_{u3}=:t^2$ and $U_{u}=U_{l}=:U$. For different parameters, we obtain the same behavior as in Ref.~\cite{Spethmann2022}. We find
\begin{align} \label{eq:H_CA}
\mathcal{H}_{\text{CA}}=\Gamma_\text{CA}^\text{2SC}\cos
\left(\frac{\varphi}{2}\right)\begin{psmallmatrix}d_{2\uparrow}^{\dagger}\\d_{2\downarrow}^{\dagger}\end{psmallmatrix}^T U^{\dagger}\begin{psmallmatrix} -d_{3\downarrow}^{\dagger}\\ d_{3\uparrow}^{\dagger}\end{psmallmatrix} + \text{H.c.}
\end{align}
Here, $\Gamma_\text{CA}^\text{2SC}=2\pi \rho_F t^2\exp[-i(\varphi_u+\varphi_l)/2]$ and we defined the superconducting phase difference $\varphi=\varphi_u-\varphi_l$. For the case of just a single superconductor as in the main text we can set the tunnel coupling to the lower superconductor to zero $t_{\alpha l}=0$, $\varphi_u=0$, and define $\Gamma_\text{CA}=\Gamma_{\text{CA},u}$, such that we obtain Eq.~\eqref{eq:H_CA_maintext} with $U_\text{so}(2\Phi_\text{so})=U_l$.

\begin{figure}[t]
	\includegraphics[width=0.49\textwidth]{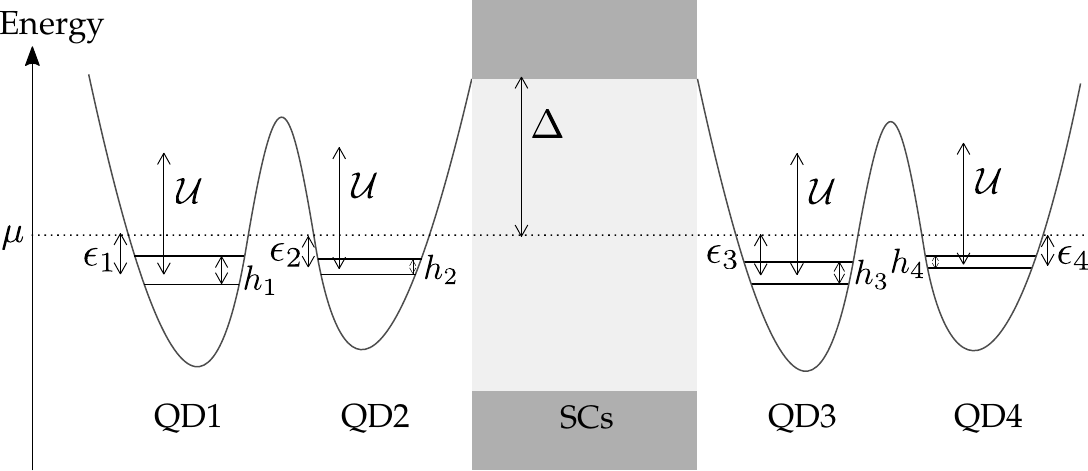}
\caption{Energy scales in our device. The superconducting gap $\Delta$ is the largest energy scale, followed by the on-site Coulomb repulsion $\mathcal{U}$, the quantum dot (QD) level energy $\epsilon_{\alpha}$ with respect to the chemical potential $\mu$ in the superconductors (SCs), and the Zeeman splitting $h_{\alpha}$. In total $0<-\epsilon_{\alpha}\pm h_{\alpha}/2<\mathcal{U}\ll\Delta$, resulting in single occupied dot states when the tunnel coupling is small. We expect our calculation to stay valid for large on-site Coulomb interaction $\mathcal{U}>\Delta$ (Appendix \ref{sec:large_Coulomb_rep}) and for intermediate Zeeman energies $h_{\alpha}\not\ll\epsilon_{\alpha},\,\mathcal{U}$ (Appendix \ref{sec:large_Zeeman_and_nonuniform_SC}).}
\label{fig:energy_scales}
\end{figure}

\subsection{Spin Hamiltonian} \label{sec:spin_Hamilt}
Following Ref.~\cite{Choi2000},  we now describe how crossed Andreev processes result in an effective spin Hamiltonian $\mathcal{H}_\text{spin}$ [Eq.~\eqref{eq:Hspin}]. We assume that single-occupied dot states are the ground state of each dot and well separated from dot states with a larger or smaller occupation. This condition requires $0<-\epsilon_{\alpha}\pm h_{\alpha}/2<\mathcal{U}$, see Fig.~\ref{fig:energy_scales}. Assuming $t_{i},\ \Gamma_{\text{CA}}^\text{(2SC)},\ h_{\alpha}\ll |\epsilon_{\alpha}|,\ \mathcal{U}$ we again apply perturbation theory.

We first review how to obtain the spin-spin interaction between dot $1$ and dot $2$ starting from our Hubbard model, neglecting the other dots \cite{Mutter2021}. This spin-spin interaction is needed for single qubit rotations of our ST qubit $i=1$. Similar to the Sec.~\ref{sec:QD_Hamilt} we first perform the basis transformation for dot $1$,
\begin{align}
\begin{pmatrix}\Tilde{d}_{1\uparrow}\\\Tilde{d}_{1\downarrow}\end{pmatrix} =U_{1} \begin{pmatrix}d_{1\uparrow}\\d_{1\downarrow}\end{pmatrix}.
\end{align}
In this frame, we have spin-conserving tunneling between dot $1$ and dot $2$,
$t_{1}\sum_{\sigma}d_{2\sigma}^{\dagger}\Tilde{d}_{1\sigma} + \text{H.c.}$ %
By assuming a large on-site repulsion $\mathcal{U}\gg t_{1}$ compared to $t_1$ and a detuning $\tilde{\epsilon}_1=\epsilon_1-\epsilon_2$ that is smaller than $\mathcal{U}$, we can reduce the six-dimensional Hilbert  space of two particles ${\{|\Tilde{\uparrow}_1\!\!\uparrow_2\rangle}$, ${|\Tilde{\uparrow}_1\!\!\downarrow_2\rangle}$, ${|\Tilde{\downarrow}_1\!\!\uparrow_2\rangle}$, ${|\Tilde{\downarrow}_1\!\downarrow_2\rangle}$, ${|\Tilde{\uparrow}_1\Tilde{\downarrow}_1\rangle}$, $|\!\!\uparrow_2\downarrow_2\rangle\}$ to the four-dimensional Hilbert space ${\{|\Tilde{\uparrow}_1\uparrow_2\rangle}$, ${|\Tilde{\uparrow}_1\downarrow_2\rangle}$, ${|\Tilde{\downarrow}_1\uparrow_2\rangle}$, ${|\Tilde{\downarrow}_1\downarrow_2\rangle\}}$ by perturbation theory [Eq. \eqref{eq:092834753048}]. The second-order contribution
 in the  small tunneling $t_1$ is 
\begin{align}
&\begin{pmatrix}
0&0&0&0\\0&-J_1/2&J_1/2&0\\0&J_1/2&-J_1/2&0\\0&0&0&0\end{pmatrix} 
= \frac{J_{1}}{4}(\Tilde{\vec{\sigma}}^1\cdot\vec{\sigma}^2 -1)\ , \\
&J_1=\frac{4t_{1}^2\mathcal{U}}{(\mathcal{U}-\tilde{\epsilon}_1)(\mathcal{U}+\tilde{\epsilon}_1)}\label{eq:J1_SM}\ .
\end{align}
Here, $\Tilde{\vec{\sigma}}^1$ ($\vec{\sigma}^2$) is a Pauli matrix describing spin $1$ ($2$) in the basis where spin $1$ is rotated by $U_1$ with respect to spin $2$, and we assumed small Zeeman fields compared to the other dot energy scales, $h_{\alpha}\ll\mathcal{U}\pm\tilde{\epsilon}_1$. 
By rotating back to the laboratory frame, we derive the anisotropic spin-spin interaction (up to an irrelevant constant)
\begin{align}\label{eq:qubit_1_exchange}
\frac{J_{1}}{4}(U^{1\dagger}\vec{\sigma}^1 U^{1})\cdot \vec{\sigma}^2 = \frac{J_1}{4} \vec{\sigma}^1 \cdot \hat{R}_1 \vec{\sigma}^2\ .
\end{align}
The $3 \times 3$ rotation matrix $\hat{R}_1$ creates a right-handed rotation of angle $\Phi_1$ and unit vector $\vec{n}_1$ related by $U_1^\dagger\vec{\sigma}^1 U_1=\hat{R}_1^{-1}\vec{\sigma}^1$ 
with $U_1=\exp(i\Phi_1\vec{n}_1\cdot\vec{\sigma}^1/2)$.
The spin-spin interaction between dots $3$ and $4$ follows analogously with the definitions $\tilde{\epsilon}_2=\epsilon_3-\epsilon_4$ and $J_2=4t_{2}^2\mathcal{U}/[(\mathcal{U}-\tilde{\epsilon}_2)(\mathcal{U}+\tilde{\epsilon}_2)]$.

Similarly, we can now evaluate the spin-spin interaction between dots $2$ and $3$ mediated by the superconductors, omitting dots $1$ and $4$. We work again in the rotated basis of dot 2, Eq.~\eqref{eq:basis_trafo_d2}. While the direct tunnel coupling between dot $1$ and $2$ (or dots $3$ and $4$) conserves the particle number, the crossed Andreev processes from the superconductor only conserve the parity. 
Therefore, we start in a basis  ${\{|\tilde{\uparrow}_2\!\!\uparrow_3\rangle}$, ${|\tilde{\uparrow}_2\!\downarrow_3\rangle}$, ${|\tilde{\downarrow}_2\!\uparrow_3\rangle}$, ${|\tilde{\downarrow}_2\!\downarrow_3\rangle}$, ${|0\rangle,~|\tilde{\uparrow}_2\!\tilde{\downarrow}_2\rangle}$, ${|\uparrow_3\!\downarrow_3\rangle}$, ${|\tilde{\uparrow}_2\tilde{\downarrow}_2\!\uparrow_3\downarrow_3\rangle\}}$ and find the effective Hamiltonian of the subspace ${\{|\tilde{\uparrow}_2\uparrow_3\rangle}$, ${|\tilde{\uparrow}_2\!\downarrow_3\rangle}$, ${|\tilde{\downarrow}_2\!\uparrow_3\rangle}$, ${|\tilde{\downarrow}_2\!\downarrow_3\rangle\}}$ by second-order perturbation theory. 
We get the second-order contribution
\begin{align}\label{eq:J_vs_SC}
&\begin{pmatrix}
0&0&0&0\\0&-\mathcal{J}(\varphi)/2&\mathcal{J}(\varphi)/2&0\\0&\mathcal{J}(\varphi)/2&-\mathcal{J}(\varphi)/2&0\\0&0&0&0\end{pmatrix}=\frac{\mathcal{J}(\varphi)}{4}(\Tilde{\vec{\sigma}}^2\cdot\vec{\sigma}^3-1)\ ,\\
&\mathcal{J}(\varphi)=-\cos^2\left(\frac{\varphi}{2}\right)\frac{4|\Gamma_\text{CA}^\text{2SC}|^2 \mathcal{U}}{(\epsilon_2+\epsilon_3)(2\mathcal{U}+\epsilon_2+\epsilon_3)}\ ,\label{eq:J_of_varphi}
\end{align}
 where $\Tilde{\vec{\sigma}}^2$ ($\vec{\sigma}^3$) is a Pauli matrix of spin $2$ ($3$) which acts in the basis where spin $2$ is rotated with respect to $3$, and where we assumed small Zeeman fields compared to the dot energy scales $h_{\alpha}\ll \mathcal{U},\ |\epsilon_{\alpha}|$. Since we are working with a simple Hubbard model and neglect, for example, interdot capacitance, the Coulomb repulsion energies $\mathcal{U}$ that enter $J_i$ [Eq.~\eqref{eq:J1_SM}] and $\mathcal{J}(\varphi)$ [Eq.~\eqref{eq:J_of_varphi}] may differ in real experiments \cite{Li2012,geyer2022two}. However, even if $J_i$ and $\mathcal{J}(\varphi)$ are renormalized, the anisotropy of the spin-spin coupling is not affected. Rotating back to the laboratory frame we derive an anisotropic spin-spin interaction similar to Eq.~\eqref{eq:qubit_1_exchange},
\begin{align}
\frac{\mathcal{J}(\varphi)}{4}(U^{\dagger}\vec{\sigma}^2 U)\cdot \vec{\sigma}^3= \frac{\mathcal{J}(\varphi)}{4} \vec{\sigma}^2 \cdot \hat{R}\vec{\sigma}^3\ .
\end{align}
As before, the rotation matrix $\hat{R}$ generates a right-handed rotation of angle $\Phi$ and unit vector $\vec{n}$ related by $U^\dagger\vec{\sigma}^2 U=\hat{R}^{-1} \vec{\sigma}^2$ 
with $U=\exp(i\Phi\vec{n}\cdot\vec{\sigma}^2/2)$.
We then find the complete Hamiltonian of our system,
\begin{align}
\label{eq:H_spin0}
\mathcal{H}_{\text{spin}}=&\frac{1}{2}\sum_{\alpha}  \vec{h}_{\alpha}\cdot\pmb{\sigma}^{\alpha} +\frac{\mathcal{J}(\varphi)}{4} \pmb{\sigma}^{2}\cdot \hat{R} \pmb{\sigma}^{3}\nonumber\\
&+\frac{ J_1}{4}\pmb{\sigma}^{1}\cdot \hat{R}_1 \pmb{\sigma}^{2}
+\frac{ J_2}{4} \pmb{\sigma}^{3}\cdot \hat{R}_2 \pmb{\sigma}^{4}\ ,
\end{align}
where we have dropped the spin-independent term $-\mathcal{J}(\varphi)/4$.
This Hamiltonian simplifies to the spin Hamiltonian in the main text by setting the rotations along the same directions and the same rotation angles $\hat{R}_1=\hat{R}_2=\hat{R}_\text{so}(\Phi_\text{so})$ and $\hat{R}=\hat{R}_\text{so}(2\Phi_\text{so})$. In addition, when we have just one superconductor, we replace $\mathcal{J}(\varphi)$ by 
 $\mathcal{J}$ as defined in Eq. \eqref{eq:couplingconstants}.

\subsection{Singlet-triplet Hamiltonian}\label{sec:ST_Hamilt}

We now calculate the ST Hamiltonian $\mathcal{H}_\text{ST}$ [Eq.~\eqref{eq:098345034542}] by a projection into the ST qubit space \cite{Li2012}. This projection is a suitable description of the qubits if the qubit states are clearly separated in energy from the other states, or if the coupling to other states is small enough. We reach this regime by considering a large Zeeman splitting $h_{\alpha} \gg J_i,\ |\mathcal{J}(\varphi)|$, which separates our ST space from states with nonzero total spin-$z$, $S^z\neq 0$.
In addition, we need a large Zeeman energy difference between dots $2$ and $3$, $|\delta h|=|h_2-h_3|\gg J_i,\ |\mathcal{J}(\varphi)|$, or we need to tune the SOI rotation to  prevent coupling to the leakage states having $S^z=0$, i.e.,  $|\uparrow_1\uparrow_2\downarrow_3\downarrow_4\rangle$  and $|\downarrow_1\downarrow_2\uparrow_3\uparrow_4\rangle$. In Appendix \ref{sec:leakage_coupling} we will explicitly calculate this SOI-dependent coupling to leakage states.
In the following, we drop the dot index $\alpha$ of the state $|\sigma_{\alpha}\rangle$ when it is clear context-wise.
 The ST subspace is spanned by the states
\begin{align} 
&\{|00\rangle,|01\rangle,|10\rangle,|11\rangle\} \text{ with} \label{eq:10928437325}\\\nonumber
& |00\rangle=|\uparrow\downarrow\uparrow\downarrow\rangle\ ,\\\nonumber\
& |01\rangle=|\uparrow\downarrow\downarrow\uparrow\rangle\ ,\\\nonumber\
& |10\rangle=|\downarrow\uparrow\uparrow\downarrow\rangle\ ,\\\nonumber\
& |11\rangle=|\downarrow\uparrow\downarrow\uparrow\rangle\ .
\end{align}
This basis choice is more practical for our calculations than the actual singlet and triplet states.
We use the following identity
\begin{align}
&\vec{\sigma}^{\alpha}\cdot \hat{R}\vec{\sigma}^{\beta} = \left(e^{-i\Phi \vec{n}\cdot\vec{\sigma}^{\alpha}/2} \vec{\sigma}^{\alpha} e^{i\Phi \vec{n}\cdot\vec{\sigma}^{\alpha}/2}\right) \cdot \vec{\sigma}^{\beta}\\\nonumber
&=\vec{\sigma}^{\alpha}\cdot\vec{\sigma}^{\beta}\cos \Phi+ (\vec{n}\cdot\vec{\sigma}^{\alpha})(\vec{n}\cdot\vec{\sigma}^{\beta})(1-\cos\Phi) \\\nonumber
&\quad- \vec{n}\cdot(\vec{\sigma}^{\alpha}\times\vec{\sigma}^{\beta}) \sin \Phi\ .
\end{align}
A projection to our basis only allows terms proportional to $\mathbb{1}$, $\sigma^{\alpha}_{z}$, $\sigma^{\alpha}_z\sigma^{\beta}_{z}$, $\sigma^{1}_{j}\sigma^{2}_{j'}$  and $\sigma^{3}_{j}\sigma^{4}_{j'}$, with $j,j'\in\{x,y\}$. 
With this consideration, we obtain the Hamiltonian (up to a constant)%
 \begin{align}\label{eq:H_ST0}
\mathcal{H}_{\text{ST}}&=\frac{1}{2}\vec{\mathcal{B}}_1\cdot \vec{\tau}^1 + \frac{1}{2}\vec{\mathcal{B}}_2 \cdot \vec{\tau}^2 + \frac{\mathcal{J}(\varphi)\gamma_{\parallel}(\Phi,\theta_0)}{4} \tau_z^1 \tau_z^2\ ,
\end{align}
with
\begin{align}
\vec{\mathcal{B}}_i&=\big( J_i \text{Re}\left[\gamma_{\perp}(\Phi_i, \theta_i)\right], J_i \text{Im}\left[\gamma_{\perp}(\Phi_i,\theta_i)\right], \delta h_i\big)^T \ , \nonumber\\
\gamma_\parallel(\Phi,\theta) &=2\sin^2(\theta)\sin^2(\Phi/2)-1\ ,  \nonumber\\ 
\gamma_\perp(\Phi,\theta) &=  \left[\cos\left(\Phi/2\right)+i\cos(\theta)\sin\left(\Phi/2\right)\right]^2\ ,%
\label{eq:gamma_and_B}
 \end{align}
 where $\vec{\tau}^1$  ($\vec{\tau}^2$) is the Pauli vector acting on the computational space of the first (second) ST qubit, spanned by the states \{$|\!\!\uparrow_1\downarrow_2\rangle, |\!\!\downarrow_1\uparrow_2\rangle$\} \big(\{$|\!\!\uparrow_3\downarrow_4\rangle$, $|\!\!\downarrow_3\uparrow_4\rangle$\}\big). The angles $\theta_i=\arccos(n_i^z)$ [$\theta_0=\arccos(n^z)$] are the angles between the Zeeman splitting direction (here the $z$ direction $\vec{e}_z$) and the spin-orbit fields $\vec{n}_i$ [$\vec{n}$]. 
 The exchange couplings $J_i$ and the Zeeman splittings $\delta h_{i}$ enable single-qubit gates, while the superconductor-mediated interaction $\mathcal{J}(\varphi)$ results in an Ising-like two-qubit interaction and can be turned on and off by the superconducting phase $\varphi$.

 Equation \eqref{eq:H_ST0} corresponds to Eq.~\eqref{eq:098345034542} where $\mathcal{J}(\varphi)$ is replaced by $\mathcal{J}$ [Eq.~\eqref{eq:couplingconstants}] in the case of only one superconductor. When we have two superconductors forming a junction, the superconducting phase difference $\varphi$ becomes a control knob to completely switch the two-qubit interaction on and off, Eq.~\eqref{eq:J_of_varphi}, suppressing crosstalk.

\section{Arbitrary direction of the Zeeman field}\label{sec:arbitr_vec_h}

In general SOI renders the $g$ tensors different in each quantum dot. Therefore, even when a homogeneous magnetic field is applied, the direction of the Zeeman field can be different for each dot.  To capture this effect, we consider the arbitrary-oriented Zeeman fields $\sum_{\alpha\sigma\sigma'}(\vec{h}_{\alpha}\cdot\vec{\sigma})_{\sigma\sigma'}d_{\alpha\sigma}^{\dagger}d_{\alpha\sigma'}/2$.  Defining $\vec{h}_{\alpha}=h_{\alpha}\hat{R}_{\mathcal{Z}\alpha} \vec{e}_Z$ with a local spin rotation $\hat{R}_{\mathcal{Z}\alpha}$ (with angle $\Phi_{\mathcal{Z}\alpha}$ and vector $\vec{n}_{\mathcal{Z}\alpha}$) one can map this general system to the system modeled by the Hamiltonian in Eqs.~\eqref{eq:HDQDsp2}, \eqref{eq:H_S_SM}, and \eqref{eq:H_T_SM}. For this a basis transformation is used,
\begin{align}
\begin{pmatrix}d_{\alpha\uparrow}\\d_{\alpha\downarrow}\end{pmatrix} \rightarrow e^{i \frac{\Phi_{\mathcal{Z}\alpha}}{2} \vec{n}_{\mathcal{Z}\alpha}\cdot \vec{\sigma}}
\begin{pmatrix}d_{\alpha\uparrow}\\d_{\alpha\downarrow}\end{pmatrix}\ ,
\end{align}
that aligns the individual Zeeman fields of the dots along the $z$ direction. This basis transformation also transforms the spin-orbit rotation matrices that now include the effects of the $g$ tensors anisotropies as
\begin{align}
&\hat{R}_{1} \rightarrow 
\hat{R}^{-1}_{\mathcal{Z}1} \hat{R}_{1}
\hat{R}_{\mathcal{Z}2}\ ,\\\nonumber
&\hat{R} \rightarrow 
\hat{R}^{-1}_{\mathcal{Z}2} \hat{R}
\hat{R}_{\mathcal{Z}3}\ ,\\\nonumber
&\hat{R}_{2} \rightarrow 
\hat{R}^{-1}_{\mathcal{Z}3} \hat{R}_{2}
\hat{R}_{\mathcal{Z}4}\ .
\end{align}
In the general case, the angles $\Phi_i, \Phi, \theta_i,\theta_0$ that enter our final result in Eq.~\eqref{eq:H_ST0} and determine the amount of leakage are those of the effective, combined rotations. Thus, $g$-anisotropies do not affect our conclusions. Also, we point out  that the anisotropy induced by the $g$ tensor difference alone is small in typical semiconductors \cite{geyer2022two} and so to a good approximation, one can simply align the magnetic field to a plane perpendicular to the SOI as discussed in the main text.

\section{Large Coulomb interaction}\label{sec:large_Coulomb_rep}

Our results, the ST Hamiltonian Eq.~\eqref{eq:H_ST0} and the leakage coupling, hold also for large Coulomb interaction $\mathcal{U}$, even when $\mathcal{U}$ exceeds the superconducting gap $\Delta$. The calculation of the crossed Andreev contribution $\mathcal{H}_\text{CA}$ is the same as for small and for large $\mathcal{U}$. In the case of $\mathcal{U}>\Delta$, for the calculation of the spin Hamiltonian, we only allow those virtual processes where the dots are occupied by zero or one particle. In that case, we find 
\begin{align}
\mathcal{J}(\varphi)=-\cos^2\left(\frac{\varphi}{2}\right)\frac{2|\Gamma_\text{CA}^{2SC}|^2}{\epsilon_2+\epsilon_3}\ ,
\end{align}
which agrees with Eq.~\eqref{eq:J_vs_SC} in the limit $\mathcal{U}\rightarrow\infty$. Thus, our theory also holds for large on-site Coulomb repulsion $\mathcal{U}$.

\section{Coupling to leakage states}\label{sec:leakage_coupling}
Here we analyze how the ST qubits couple to the leakage states $|\uparrow\uparrow\downarrow\downarrow\rangle$ and $|\downarrow\downarrow\uparrow\uparrow\rangle$ \cite{Klinovaja2012}. To achieve this goal, we project the spin Hamiltonian $\mathcal{H}_\text{spin}$ [Eq.~\eqref{eq:H_spin0}] onto the full six-dimensional subspace with zero spin-$z$ $S^z=0$ using the same steps as in the calculation of $\mathcal{H}_\text{ST}$ [Eq.~\eqref{eq:H_ST0}].
We find (up to a constant) 
\begin{widetext}
\begin{align}
\label{eq:H_leakage}
&\mathcal{H}_{6\times 6}=\mathcal{H}_\text{ST} + \Bigg[ \frac{\mathcal{J}(\varphi) \gamma_{\perp} (\Phi,\theta_0)}{2} \bigg( |\uparrow\downarrow\uparrow\downarrow\rangle \langle\uparrow\uparrow\downarrow\downarrow|  %
+|\downarrow\downarrow\uparrow\uparrow\rangle \langle\downarrow\uparrow\downarrow\uparrow|\bigg) + \text{H.c.}\Bigg]%
+E^{leak}_+ |\uparrow\uparrow\downarrow\downarrow\rangle\langle \uparrow\uparrow\downarrow\downarrow|%
+ E^{leak}_- \textstyle |\downarrow\downarrow\uparrow\uparrow\rangle \langle \downarrow\downarrow\uparrow\uparrow|,\nonumber\\
 &E^{leak}_{\pm}= \pm \frac{h_1+h_2-h_3-h_4}{2}+\sum_i \frac{J_i\gamma_{\parallel}(\Phi_i,\theta_i)}{2} + \frac{\mathcal{J}(\varphi)\gamma_{\parallel}(\Phi,\theta_0)}{4}\ . %
\end{align}
\end{widetext}
The coupling to the leakage states is given by $\mathcal{J}(\varphi)\gamma_{\perp}(\Phi,\theta_0)/2$. The rotation angle $\Phi$ is related to the $\Phi_{\text{so}}$ in the main text by $\Phi=2\Phi_\text{so}$ and corresponds to the total angle a particle would rotate when tunneling from dot $2$ to dot $3$. Similarly, $\theta_0$ corresponds to $\theta$ in the main text. For $\Phi=2\Phi_\text{so}=\pi$ and $\theta_0=\theta=\pi/2$ the leakage coupling becomes zero while the two-qubit interaction $\propto\gamma_{\parallel}(\Phi,\theta_0)$ becomes maximal [Eq.~\eqref{eq:gamma_and_B}].

 We note that leakage is not a problem of single-qubit rotations because for an individual qubit the leakage states (the triplets $|T_+\rangle=|\uparrow\uparrow\rangle$ and $|T_-\rangle=|\downarrow\downarrow\rangle$) are separated from the computational states by the large global magnetic field. We also note that it is not required to adjust $\Phi_i$, $\Phi$, $\theta_0$, or $\theta_i$ in between quantum operations. They need to be calibrated only once at the beginning. During this calibration, $\Phi_i$ and $\theta_i$ should be chosen such that $\gamma_{\perp}(\Phi_i,\theta_i)$ is not small, as this allows one to perform single-qubit rotations by a modulation of $J_i$. 
\\\\

\onecolumngrid

\section{Large Zeeman fields and nonuniform coupling to the superconductors}\label{sec:large_Zeeman_and_nonuniform_SC}

The calculation in the main text and in Appendix \ref{sec:spin_Hamilt} assumed that the Zeeman energies were much smaller than the other dot energies, $h_{\alpha}\ll|\epsilon_{\alpha}|,\mathcal{U}$. Further, we assumed that the tunnel couplings to the lower superconductors were the same as the tunnel couplings for the upper superconductors, $t_{l2}t_{l3}=t_{u2}t_{u3}$, and that the spin-orbit induced spin rotations were also the same for both superconductors, $U_l=U_r$. Those assumptions were needed to derive the spin Hamiltonian in Eq. \eqref{eq:H_spin0}.
In this section we show that our results of vanishing leakage and crosstalk hold even if the Zeeman fields are of the same order of magnitude as the dot energy scales, $h_{\alpha} \lesssim|\epsilon_{\alpha}|,\mathcal{U}$, and for deviations from $t_{l2}t_{l3}=t_{u2}t_{u3}$ and $U_l=U_r$ up to linear order. Here, we directly derive the effective singlet-triplet Hamiltonian.

We start from Eq. \eqref{eq:H_CA_general} and assume that coupling parameters $|\Gamma_{\text{CA,j}}|$ deviate only a little from a common, $j$-independent value $\bar{\Gamma}_{\text{CA}}$. Similarly, we assume that the spin-orbit rotations $U_j$ are approximate  $\pi$ rotations around the $y$ axis [this corresponds to $\Phi=\pi$, $\theta_0=\pi/2$ in Eq. \eqref{eq:H_ST0}, or $\Phi_\text{so}=\pi/2$ and $\theta=\pi/2$ in the main text]. 
Thus, we have 
\begin{align}
\Gamma_{\text{CA,j}}&=\exp(-i\varphi_j)\bar{\Gamma}_{\text{CA}}(1+\delta_{\text{CA,j}}) ,\\
U_j&=\textstyle \exp\left(i\frac{\delta \Phi_j}{2}\vec{\delta n}_j \cdot \vec{\sigma}\right)\, i\sigma_y\approx \left(1+i\frac{\delta \Phi_j}{2}\vec{\delta n}_j \cdot \vec{\sigma}\right)i\sigma_y ,
\end{align}
with some unit vectors $\vec{\delta n}_j$, small parameters $\delta_{\text{CA,j}}$ and $\delta \Phi_j$, and Pauli vector $\vec{\sigma}=(\sigma_x\, \sigma_y\,\sigma_z)^T$. We chose the $y$-axis as the direction of the spin-orbit interaction (meaning that $U_j\approx i\sigma_y$),  but any other direction perpendicular to the Zeeman field will work equally well. 
To linear order in $\delta_{\text{CA,j}}$ and $\delta\Phi_j$, the crossed Andreev term $\mathcal{H}_{\text{CA}}$, Eq.~\eqref{eq:H_CA_general}, is now
\begin{align}
\mathcal{H}_{\text{CA}}\approx \bar{\Gamma}_{\text{CA}}e^{-i(\varphi_u+\varphi_l)/2}(-1)&\Bigl\{ \left(2\cos(\varphi/2)+e^{-i\varphi/2}\delta_\text{CA,u}+e^{i\varphi/2}\delta_\text{CA,l}\right)\left(d_{2\uparrow}^{\dagger}d_{3\uparrow}^{\dagger}+d_{2\downarrow}^{\dagger}d_{3\downarrow}^{\dagger}\right)\nonumber\\\nonumber
&+\textstyle e^{-i\varphi/2}\frac{\delta\Phi_u}{2}\left[(d_{2\uparrow}^{\dagger}d_{3\uparrow}^{\dagger}-d_{2\downarrow}^{\dagger}d_{3\downarrow}^{\dagger})i\delta n_u^z + d_{2\uparrow}^{\dagger}d_{3\downarrow}^{\dagger}(i\delta n_u^x-\delta n_u^y) +d_{2\downarrow}^{\dagger}d_{3\uparrow}^{\dagger}(i\delta n_u^x+\delta n_u^y)\right]\\\nonumber
&+\textstyle e^{i\varphi/2}\frac{\delta\Phi_l}{2}\left[(d_{2\uparrow}^{\dagger}d_{3\uparrow}^{\dagger}-d_{2\downarrow}^{\dagger}d_{3\downarrow}^{\dagger})i\delta n_l^z + d_{2\uparrow}^{\dagger}d_{3\downarrow}^{\dagger}(i\delta n_l^x-\delta n_l^y) +d_{2\downarrow}^{\dagger}d_{3\uparrow}^{\dagger}(i\delta n_l^x+\delta n_l^y)\right]\Bigr\}\\
+H.c.\qquad\qquad\qquad\quad&
\end{align}
Next we assume that all dots are occupied by one electron, $0<-\epsilon_{\alpha}\pm h_{\alpha}<\mathcal{U}$, and that the crossed Andreev coupling is much smaller than the Zeeman energy and other dot energy scales $\bar{\Gamma}_{\text{CA}}\ll h_{\alpha},|\epsilon_{\alpha}|,\mathcal{U}$.

We then directly construct an effective Hamiltonian $\mathcal{H}_\text{ST}^\text{int}$ by second-order perturbation theory [Eq.~\eqref{eq:092834753048}] in $\mathcal{H}_\text{CA}$. As before, this effective Hamiltonian acts in the space with total spin $z$--component $S^z=0$, which is spanned by the singlet-triplet basis states, Eq.~\eqref{eq:10928437325}, and the two leakage states, $|\uparrow_1\uparrow_2\downarrow_3\downarrow_4\rangle$ and $|\downarrow_1\downarrow_2\uparrow_3\uparrow_4\rangle$. %
Since $\mathcal{H}_\text{CA}$ does not directly affect dot 1 or dot 4, the only nonzero matrix elements of $\mathcal{H}_\text{ST}^\text{int}$ are $\langle\uparrow_2\uparrow_3|\mathcal{H}_\text{ST}^\text{int}|\uparrow_2\uparrow_3\rangle$, $\langle \downarrow_2\downarrow_3|\mathcal{H}_\text{ST}^\text{int}|\downarrow_2\downarrow_3\rangle$, $\langle \uparrow_2\downarrow_3|\mathcal{H}_\text{ST}^\text{int}|\uparrow_2\downarrow_3\rangle$, $\langle \downarrow_2\uparrow_3|\mathcal{H}_\text{ST}^\text{int}|\downarrow_2\uparrow_3\rangle$, $\langle \uparrow_2\downarrow_3|\mathcal{H}_\text{ST}^\text{int}|\downarrow_2\uparrow_3\rangle$ and $\langle \downarrow_2\uparrow_3|\mathcal{H}_\text{ST}^\text{int}|\uparrow_2\downarrow_3\rangle$. Since those matrix elements hold for any spin on dot 1 and 4, we omitted dot 1 and 4 in the notation. %
We find easily that to linear order in $\delta \Phi_j$, the last four terms are zero, thus, coupling to leakage states of the singlet-triplet qubits  will be zero. The first two terms are
\begin{align}
\mathcal{J}_{\uparrow\uparrow}(\varphi)=\langle \uparrow_2\uparrow_3|\mathcal{H}_\text{ST}^\text{int}|\uparrow_2\uparrow_3\rangle=&\textstyle2\cos(\frac{\varphi}{2})\frac{\bar{\Gamma}_\text{CA}^2 }{\epsilon_2+\epsilon_3+h_2+h_3}
\left[(2+\delta_\text{CA,u}+\delta_\text{CA,l})\cos\frac{\varphi}{2} +\left(\frac{\delta\Phi_u}{2}\delta n_u^z-\frac{\delta\Phi_l}{2}\delta n_l^z\right)\sin\frac{\varphi}{2}\right]\\\nonumber
&-\textstyle2\cos(\frac{\varphi}{2})\frac{\bar{\Gamma}_\text{CA}^2}{\epsilon_2+\epsilon_3-h_2-h_3+2\mathcal{U}}\left[(2+\delta_\text{CA,u}+\delta_\text{CA,l})\cos\frac{\varphi}{2} -\left(\frac{\delta\Phi_u}{2}\delta n_u^z-\frac{\delta\Phi_l}{2}\delta n_l^z\right)\sin\frac{\varphi}{2}\right] ,\\
\mathcal{J}_{\downarrow\downarrow}(\varphi)=\langle \downarrow_2\downarrow_3|\mathcal{H}_\text{ST}^\text{int}|\downarrow_2\downarrow_3\rangle=&\textstyle2\cos(\frac{\varphi}{2})\frac{\bar{\Gamma}_\text{CA}^2}{\epsilon_2+\epsilon_3-h_2-h_3}\left[(2+\delta_\text{CA,u}+\delta_\text{CA,l})\cos\frac{\varphi}{2} -\left(\frac{\delta\Phi_u}{2}\delta n_u^z-\frac{\delta\Phi_l}{2}\delta n_l^z\right)\sin\frac{\varphi}{2}\right]\\ \nonumber
&-2\textstyle\cos(\frac{\varphi}{2})\frac{\bar{\Gamma}_\text{CA}^2}{\epsilon_2+\epsilon_3
+h_2+h_3+2\mathcal{U}}\left[(2+\delta_\text{CA,u}+\delta_\text{CA,l})\cos\frac{\varphi}{2} +\left(\frac{\delta\Phi_u}{2}\delta n_u^z-\frac{\delta\Phi_l}{2}\delta n_l^z\right)\sin\frac{\varphi}{2}\right] .
\end{align}
The $\mathcal{J}_{\uparrow\uparrow}(\varphi)$ and $\mathcal{J}_{\downarrow\downarrow}(\varphi)$ determine the two-qubit interaction and since they are proportional to $\cos(\varphi/2)$ the interaction will be zero for $\varphi=\pi$. Therefore, crosstalk also vanishes at this order in perturbation theory.
For completeness, the full singlet-triplet Hamiltonian $\mathcal{H}_\text{ST,full}$ is, including the single-qubit terms,
\begin{align}
    \mathcal{H}_{\text{ST,full}}=\frac{1}{2}\vec{\mathcal{B}}_1\cdot \vec{\tau}^1 + \frac{1}{2}\vec{\mathcal{B}}_2 \cdot \vec{\tau}^2 +
    \frac{1}{4}[\mathcal{J}_{\uparrow\uparrow}(\varphi)-\mathcal{J}_{\downarrow\downarrow}(\varphi)](\tau^1_z-\tau^2_z) -
    \frac{1}{4}\left[\mathcal{J}_{\uparrow\uparrow}(\varphi)+\mathcal{J}_{\downarrow\downarrow}(\varphi)\right] (\tau_z^1 \tau_z^2\ -1).
\end{align}
We obtained the single-qubit terms in an equivalent way by second-order perturbation theory with respect to the tunneling within the double quantum dots, but this time only requiring small Zeeman field differences $|\delta h_1|=|h_1-h_2|\ll|\mathcal{U}\pm\tilde{\epsilon}_1|$ and $|\delta h_2|=|h_3-h_4|\ll|\mathcal{U}\pm\tilde{\epsilon}_2|$ instead of small total Zeeman fields $h_\alpha$.
The additional term $\propto(\mathcal{J}_{\uparrow\uparrow}(\varphi)-\mathcal{J}_{\downarrow\downarrow}(\varphi))$ in $\mathcal{H}_{\text{ST,full}}$ only weakly renormalizes the single qubit energies. The result is consistent with Eq. \eqref{eq:H_ST0} in the limit  $h_\alpha\ll \mathcal{U},|\epsilon_{\alpha}|$, $\delta_\text{CA,j}\rightarrow 0$, and $\delta\Phi_j\rightarrow 0$. %
In conclusion, we emphasize that our proposal to reduce leakage and crosstalk in hybrid semiconducting-superconducting singlet-triplet qubits does not require any fine-tuning of parameters, including the coupling parameters to the superconductor.

\section{Higher order corrections}\label{sec:correct_leak}
\subsection{Leakage corrections}
Here, we present the coupling of the ST states to the leakage states in second-order in perturbation theory [Eq.~\eqref{eq:092834753048}], denoted by $\mathcal{H}_{6\times 6}^{(2)}$. More precisely, we start from Eq. \eqref{eq:H_spin0} and take the second order perturbation with respect to the small coupling parameters $J_i$, $\mathcal{J}(\varphi)$, instead of just the projection. We find the following terms: 
\begin{align}
&\langle\uparrow\downarrow\uparrow\downarrow \textstyle |\mathcal{H}_{6\times 6}^{(2)}|\uparrow\uparrow\downarrow\downarrow \rangle = -(\langle\downarrow\uparrow\downarrow\uparrow \textstyle |\mathcal{H}_{6\times 6}^{(2)}|\downarrow\downarrow \uparrow\uparrow\rangle)^* =
\textstyle\frac{1}{16}(\frac{1}{h_3}+\frac{1}{h_2})(\gamma_1(-\Phi_1)\beta_+ +\gamma_2(\Phi_2)\beta_- )\ , \\ \nonumber
&\langle\uparrow\downarrow\downarrow\uparrow \textstyle |\mathcal{H}_{6\times 6}^{(2)}|\uparrow\uparrow\downarrow\downarrow \rangle = -(\langle\downarrow\uparrow\uparrow \downarrow\textstyle |\mathcal{H}_{6\times 6}^{(2)}|\downarrow\downarrow \uparrow\uparrow\rangle)^* =
\textstyle -\frac{1}{8}\left(\frac{1}{h_2+h_3}+\frac{1}{h_3+h_4}\right) \mathcal{J}(\varphi) J_{2}(n^x+in^y)^2(n_{2}^x-in_2^y)^2 \sin^2{\Phi}\sin^2{\Phi_2}\ , \\ \nonumber
&\langle\downarrow\uparrow \uparrow\downarrow\textstyle |\mathcal{H}_{6\times 6}^{(2)}|\uparrow\uparrow\downarrow\downarrow \rangle = -(\langle\uparrow \downarrow\downarrow\uparrow\textstyle |\mathcal{H}_{6\times 6}^{(2)}|\downarrow\downarrow \uparrow\uparrow\rangle)^* =
\textstyle \frac{1}{8} (\frac{1}{h_1+h_2}+\frac{1}{h_2+h_3}) \mathcal{J}(\varphi) J_{1}(n^x-in^y)^2 (n_{1}^x+in_{1}^y)^2 \sin^2{\Phi}\sin^2{\Phi_2}\ ,  \\ \nonumber
&\langle\downarrow\uparrow\downarrow\uparrow\textstyle |\mathcal{H}_{6\times 6}^{(2)}|\uparrow\uparrow\downarrow\downarrow \rangle =
-\langle \uparrow\downarrow\uparrow\downarrow\textstyle |\mathcal{H}_{6\times 6}^{(2)}|\downarrow\downarrow \uparrow\uparrow\rangle^* =0\ ,\\ \nonumber
&\gamma_j(\Phi_j)=J_{j}(n_{j}^x+in_j^y)\sin{\Phi_j}(\cos{\Phi_j}+in_{j}^z \sin{\Phi_j})\ ,\\ \nonumber
&\beta_{\pm}= \mathcal{J}(\varphi)(n^x\pm in^y)\sin{\Phi}(\cos{\Phi}+in \sin{\Phi})\ .
\end{align}
We conclude two points. First, we can suppress leakage even at second order by setting $\Phi_1=\Phi_2=0$. Second, the results are inversely proportional to the large global Zeeman energy. Thus, even without setting $\Phi_1=\Phi_2=0$, the second-order corrections will be small, as small as the leakage to states with nonzero spin.

\subsection{Interactions corrections}\label{sec:correct_interact}
To the lowest (zeroth) order in perturbation theory for small $J_i$, $\mathcal{J}(\varphi)$, our two-qubit interaction was given  by $\mathcal{J}(\varphi)\gamma_{\parallel}(\Phi,\theta_0)\tau^1_z\tau^2_z/4$, Eq.~\eqref{eq:H_ST0}. Including terms to the second order ($\mathcal{H}_\text{int}^{(2)}$) results in 
\begin{align}
&\quad \mathcal{H}_\text{int}^{(2)}= \frac{J_{xz}^{(2)}}{4}\tau_x^1\tau_z^2 +\frac{J_{yz}^{(2)}}{4}\tau_y^1\tau_z^2 +\frac{J_{zx}^{(2)}}{4}\tau_z^1\tau_x^2+\frac{J_{zy}^{(2)}}{4}\tau_z^1\tau_y^2\ ,
\\ \nonumber
\qquad J_{xz}^{(2)}&=\textstyle -\frac{1}{2}(\frac{1}{h_1}+\frac{1}{h_2})J_{1}\mathcal{J}(\varphi) \sin\frac{\Phi_{1}}{2}\sin\frac{\Phi}{2}\left( C_\text{mix}^1\cos\frac{\Phi_{1}}{2} + S_\text{mix}^1 n_{1}^z\sin\frac{\Phi_{1}}{2} \right)\ ,\\ \nonumber
J_{yz}^{(2)}&=\textstyle -\frac{1}{2}(\frac{1}{h_1}+\frac{1}{h_2}) J_{1}\mathcal{J}(\varphi) \sin\frac{\Phi_{1}}{2} \sin\frac{\Phi}{2}\left(C_\text{mix}^1 n_{1}^z\sin\frac{\Phi_{1}}{2} -S_\text{mix}^1\cos\frac{\Phi_{1}}{2} \right)\ ,\\ \nonumber
\qquad J_{zx}^{(2)}&=\textstyle \frac{1}{2}(\frac{1}{h_3}+\frac{1}{h_4})J_{2}\mathcal{J}(\varphi) \sin\frac{\Phi_{2}}{2}\sin\frac{\Phi}{2}\left( C_\text{mix}^2\cos\frac{\Phi_{2}}{2} - S_\text{mix}^2 n_{2}^z\sin\frac{\Phi_{2}}{2} \right)\ ,\\ \nonumber
J_{zy}^{(2)}&=\textstyle -\frac{1}{2}(\frac{1}{h_3}+\frac{1}{h_4}) J_{2}\mathcal{J}(\varphi) \sin\frac{\Phi_{2}}{2} \sin\frac{\Phi}{2}\left(-C_\text{mix}^2 n_{2}^z\sin\frac{\Phi_{2}}{2} -S_\text{mix}^2\cos\frac{\Phi_{2}}{2} \right)\ ,\\ \nonumber
C_\text{mix}^1&=\textstyle\left(n_{1}^x n^x+n_{1}^y n^y\right)\cos\frac{\Phi}{2}-
\left(n_{1}^xn^y-n_{1}^yn^x\right)n^z\sin\frac{\Phi}{2}\ ,\\ \nonumber
S_\text{mix}^1&=\textstyle \left(n_{1}^xn^y-n_{1}^yn^x\right)\cos\frac{\Phi}{2}+
\left(n_{1}^xn^x+n_{1}^yn^y\right)n^z\sin\frac{\Phi}{2}\ ,\\\nonumber
C_\text{mix}^2&=\textstyle\left(n_{2}^x n^x+n_{2}^y n^y\right)\cos\frac{\Phi}{2}+
\left(n_{2}^xn^y-n_{2}^yn^x\right)n^z\sin\frac{\Phi}{2}\ ,\\ \nonumber
S_\text{mix}^2&=\textstyle \left(n_{2}^xn^y-n_{2}^yn^x\right)\cos\frac{\Phi}{2}-
\left(n_{2}^xn^x+n_{2}^yn^y\right)n^z\sin\frac{\Phi}{2}\ . 
\end{align}
We observe that by operating the system at our sweet spot with  $\Phi=\pi$ and $\theta_0=\pi/2$ (thus, $n^z=0$) the interaction remains Ising-like not only at lowest order but also at higher orders.

\section{Details of leakage calculations in Fig. 2(d)}\label{sec:leakage_plot_specifications}

Here we provide more details of calculating the leakage as a function of time in Fig. \ref{fig:423049850445989}(d). 
We define leakage $L$ as the probability that an initial state $|\psi_i\rangle$ leaves the computational space after time evolution with time $t$, maximized over the four basis states given in Eq. \eqref{eq:10928437325}:
\begin{align}
    L=\max_{i}\sum_j |\langle \psi_j|\exp(-itH_{\text{spin}}/\hbar)|\psi_i\rangle|^2 .
\end{align}
Here, $H_\text{spin}$ is the 16-dimensional Hamiltonian in Eq.~\eqref{eq:Hspin} for the values given in the caption of Fig.~\ref{fig:423049850445989}. The initial states $|\psi_i\rangle$ are elements of the four basis states defined in Eq. \eqref{eq:10928437325} and the states $|\psi_j\rangle$ are the other 12 basis states in the space of $H_{\text{spin}}$ that are not in the ST subspace.

\end{document}